\documentclass[prb,twocolumn,floatfix]{revtex4-1}
\usepackage[bottom]{footmisc}
\usepackage[T1]{fontenc}
\usepackage[latin2]{inputenc}
\usepackage{bm}
\usepackage{graphicx}

\begin{document}

\title{Quantum Dot--Ring Nanostructure -- a~Comparison of Different Approaches}
\author{Iwona Janus-Zygmunt}
\author{Barbara K\k{e}dzierska}
\author{Anna Gorczyca-Goraj}
\author{El\.zbieta~Zipper}
\author{Maciej M. Ma\'{s}ka}
\email[E-mail me at: ]{maciej.maska@phys.us.edu.pl}
\affiliation{Department of Theoretical Physics, Institute of Physics,
  Uniwersytecka 4, 40-881~Katowice, Poland}

\begin{abstract}
  It has been recently shown that a nanostructure composed of a quantum dot surrounded by a
  quantum ring possesses a set of very unique characteristics that make it a good candidate
  for future nanoelectronic devices. Its main advantage is the ability to easily tune
  transport properties on demand by so called ``wave function engineering''. In practice,
  the distribution of the electron wave function in the nanostructure can be controlled
  by, e.g., electrical gating. In order to predict some particular properties of the system
  one has to know the exact wave functions for different shapes of the confining potential
  that defines the structure. 

  In this paper we compare three different methods that can be used to determine the
  energy spectrum, electron wave functions and transport properties of the system under
  investigation. In the first approach we utilize the cylindrical symmetry of the
  confining potential and solve only the radial part of the Schr\"odinger equation; in
  the second approach we discretize the Schr\"odinger equation in two dimensions and
  find the eigenstates with the help of the Lancz\"os method; in the third approach we
  use package Kwant to solve a tight--binding approximation of the original system. To
  study the transport properties in all these approaches we calculate microscopically
  the strength of the coupling between the nanosystem and leads. In the first two
  approaches we use the Bardeen method, in the third one calculations are performed
  with the help of package Kwant.
\end{abstract}

\maketitle

In the recent years advances in nanofabrication have led to a growing interest in complex
semiconducting  nanostructures \cite{hans}. Such  complex systems are highly controllable
objects. By utilizing their peculiar properties one can design devices with additional
functionalities.

In this paper we consider a quasi two dimensional nanostructure in the form of a quantum dot
(QD) surrounded by a quantum ring (QR), named afterwords a {\bf d}ot-{\bf r}ing {\bf n}anostructure
(DRN). It has been shown \cite{zipper,MK0,MK1,MK2,zkm,Zeng} that by changing the confinement potential,
e.g., by electrical gating, one can change many physical characteristics such as spin relaxation,
optical absorption and conduction. All these features are strongly related to the spatial
distribution of the electron wave functions in a DRN.

In particular, conduction through a DRN depends crucially on the coupling strength of its states
to the leads, which is largely dependent on the position of the electron wave functions: states
located in a quantum dot (quantum ring) are weakly (strongly) coupled. By changing the confinement
potential this distribution can be modified so that the ground and excited states move over between
the inner (dot) and the outer (ring) part of the DRN.

Transport through a DRN separated from the source and drain leads by the tunneling barriers can be
studied by various model calculations. In this paper we want to compare results of a few different
approaches. In the first one we assume a weak coupling between the DRN and leads. It means that the
electron states in the DRN and in the leads do not change when the leads are attached to the
nanosystem. They can be calculated separately and then used to determine the tunneling rates for
tunneling between the DRN and the leads. In this case we use the Bardeen approach \cite{Bardeen,
Tersoff,Tersoff1,Chen,Chen1}
to calculate microscopically the coupling strength. The wave functions of an electron in the DRN
are calculated by solving numerically the time--independent Schr\"odinger equation. To this end,
we apply two methods: 
\begin{enumerate}
\item we exploit the cylindrical symmetry of the confining potential and numerically solve the
  equation for the radial part of the electron wave function,
\item we discretize the Schr\"odinger equation in two dimensions and then use the Lancz\"os
  method to find the low--lying eigenstates.
\end{enumerate}
As expected, for a sufficiently fine discretization the energy spectrum and the shape of the wave
functions obtained in both approaches are very similar in the physically relevant regime
\cite{Boykin}. As a result, also the transport properties in both cases are similar.

In the second approach we release the requirement of a weak coupling between the DRN and the leads.
Then, when the leads are attached the wave functions change their original shape and the Bardeen
approach cannot be used any more. Instead, we use a wave-function based approach to compute
transport properties in a tight-binding system. Namely, we perform calculations with the help of
an open-source software package Kwant \cite{Kwant}. It allows one to define the nanosystem together
with the leads, so in this approach we are able to control the coupling. By comparing results of
both methods one can determine the range of coupling strengths for which the Bardeen approximation
is valid.

Within the framework of both methods we calculate the energy spectra, spatial distributions of the
ground and excited state wave functions and transport characteristics for different shapes of the
DRN confining potential.

\section{Dot--ring nanostructure}\label{sec2}

\begin{figure}[h]
\centerline{\includegraphics[width=\linewidth]{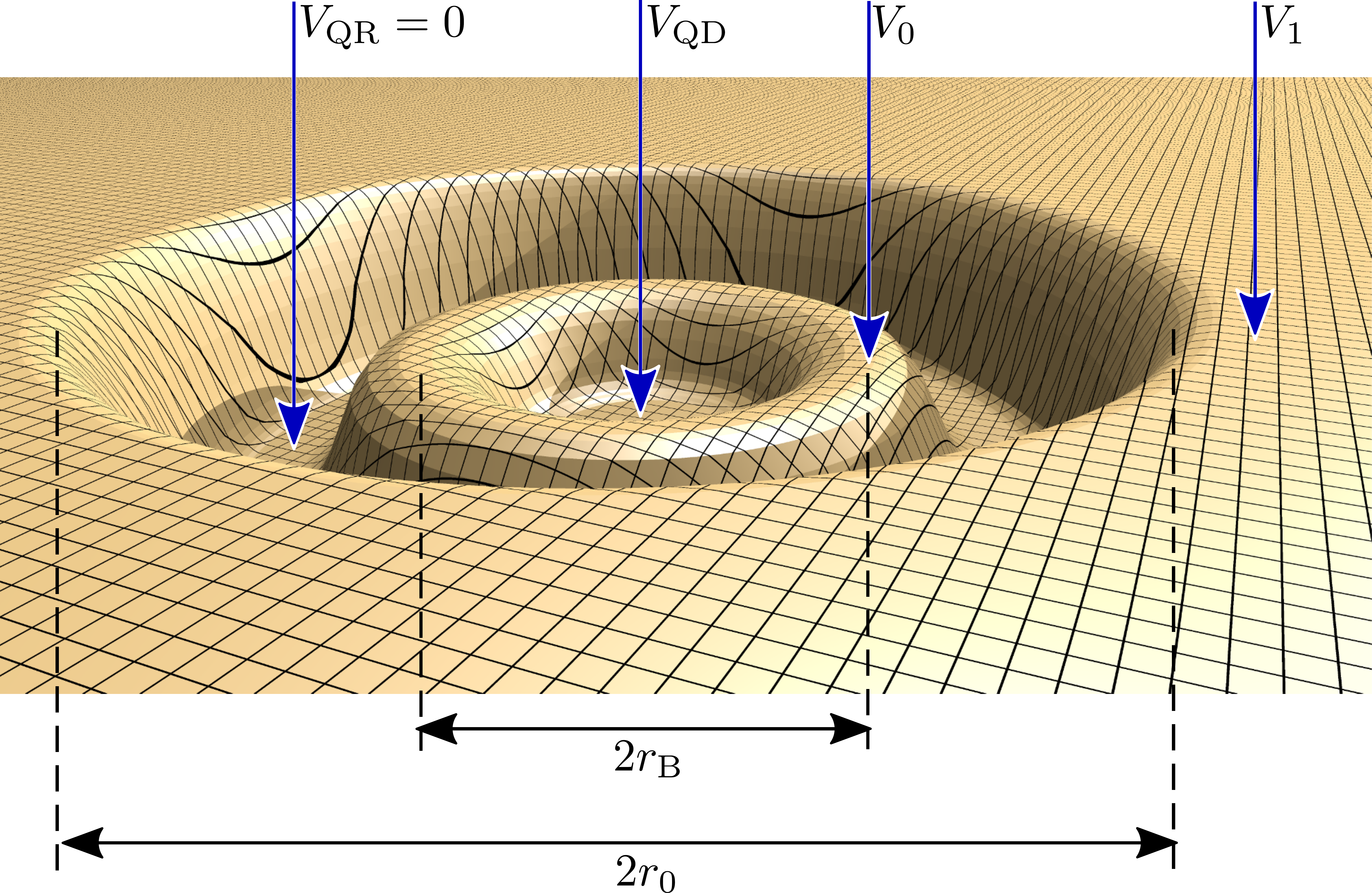}}
\caption{Illustration of the confining potential of a DRN. $r_0$ denotes the radius of the
  nanostructure, $r_{\rm B}$ denotes the radius of the potential barrier between QD and QR,
  $V_{\rm QR},\:V_{\rm QD},\:V_0$, and $V_1$ denote the potential of the QR part (which we assume to
  be the reference energy, $V_{\rm QR}=0$), QD part, the height of the potential barrier and the
  value of the potential outside the DRN, respectively.\label{fig0}} 
\end{figure}

We consider a two dimensional, circularly symmetric dot--ring nanostructure defined by a confinement
potential $V(r)$. Such a structure which conserves the circular symmetry \cite{peeters} has been
recently fabricated with the help of the droplet epitaxy method \cite{somaschini,somaschini1}. 
The DRN is composed of a quantum dot surrounded by a quantum ring with a potential barrier $V_0$ between
these parts. The height of the barrier is so that it allows for the electron tunneling between the dot
and ring parts of the DRN. The overall view of the DRN confining potential with explanation of
symbols used throughout the text is presented in Fig.~\ref{fig0}.

The confining potential is modeled by the function V(r) given by
\begin{equation}
    V(r)=\left\{\begin{array}{l} 
    V_{\rm QD} +\left(V_0-V_{\rm QD}\right)\exp\left[-0.018(r-r_{\rm B})^2\right] \\[1pt]
    \hfill\mbox{for } r< r_{\rm B}, \\[15pt]
    V_1\left\{1-\exp\left[-(r/r_0)^{30}\right]\right\}  \\[4pt]
    +V_0 \exp\left[-0.018(r-r_{\rm B})^2\right] \hfill \mbox{for } r>r_{\rm B}.
    \end{array}\right.
    \label{potential}
\end{equation}
Throughout this paper we assume the radius of the DRN $r_0=60$ nm, the position of the internal barrier
$r_{\rm B}=\frac{1}{2}r_0=30$ nm and the height of the external barrier $V_1 =90$ meV. The bottom of
the QR potential is zero, whereas the bottom of the QD potential and the height of the internal
barrier are parameters which are varied. In real experiments $V_{\rm QD}$ can be controlled by
applying voltage to a gate located close to the center of the DRN. By changing $V_{\rm QD}$ one
can modify the distribution of the electron wave function. For example, if $V_{\rm QD} \ll V_{\rm QR}$
the ground state is located in the QD, whereas for $V_{\rm QD} \gg V_{\rm QR}$ the ground state is in
the QR. Such manipulations can drastically affect the conductance through the ground state: on the
one hand, if the wave function is in QD, the overlap with states in leads is negligible and
the conductance is close to zero. On the other hand, if the wave function is in QR, its overlap
with the lead states is much larger and so is the conductance.

The energy spectrum of the original ``continuous'' system, i.e., before any discretization, consists
of a set of discrete energies $E_{nl}$ due to radial motion with radial quantum numbers
$n=0,1,2,\ldots$, and rotational motion with angular momentum quantum numbers
$l=0,\pm 1,\pm 2\ldots$. The single particle orbital  wave function is of the form 
\begin{equation}
\Psi_{nl}({\bm r}) = R_{nl}\left(r\right)\exp\left(i l \phi \right),
\label{eq_psi_nl}
\end{equation}
with the radial part $R_{nl}(r)$.

\section{Distribution of electron wave functions in a dot--ring nanostructure}
The energy spectra and wave functions have been calculated in three different models of the DRN:
\begin{enumerate}
\item A ``continuous'' system that conserves the cylindrical symmetry, without leads. The Schr\"odinger
  equation is solved only for the radial part of the wave function.
\item The Schr\"odinger equation is discretized in two dimensions. Then, the Lancz\"os method is used to
  find the ground and low--lying excited states. No leads.
\item The tight--binding version of the Hamiltonian with attached two leads is solved with the help
  of the Kwant package.
\end{enumerate}

\begin{figure}[h]
  \centerline{a)\ \includegraphics[height=2.8cm]{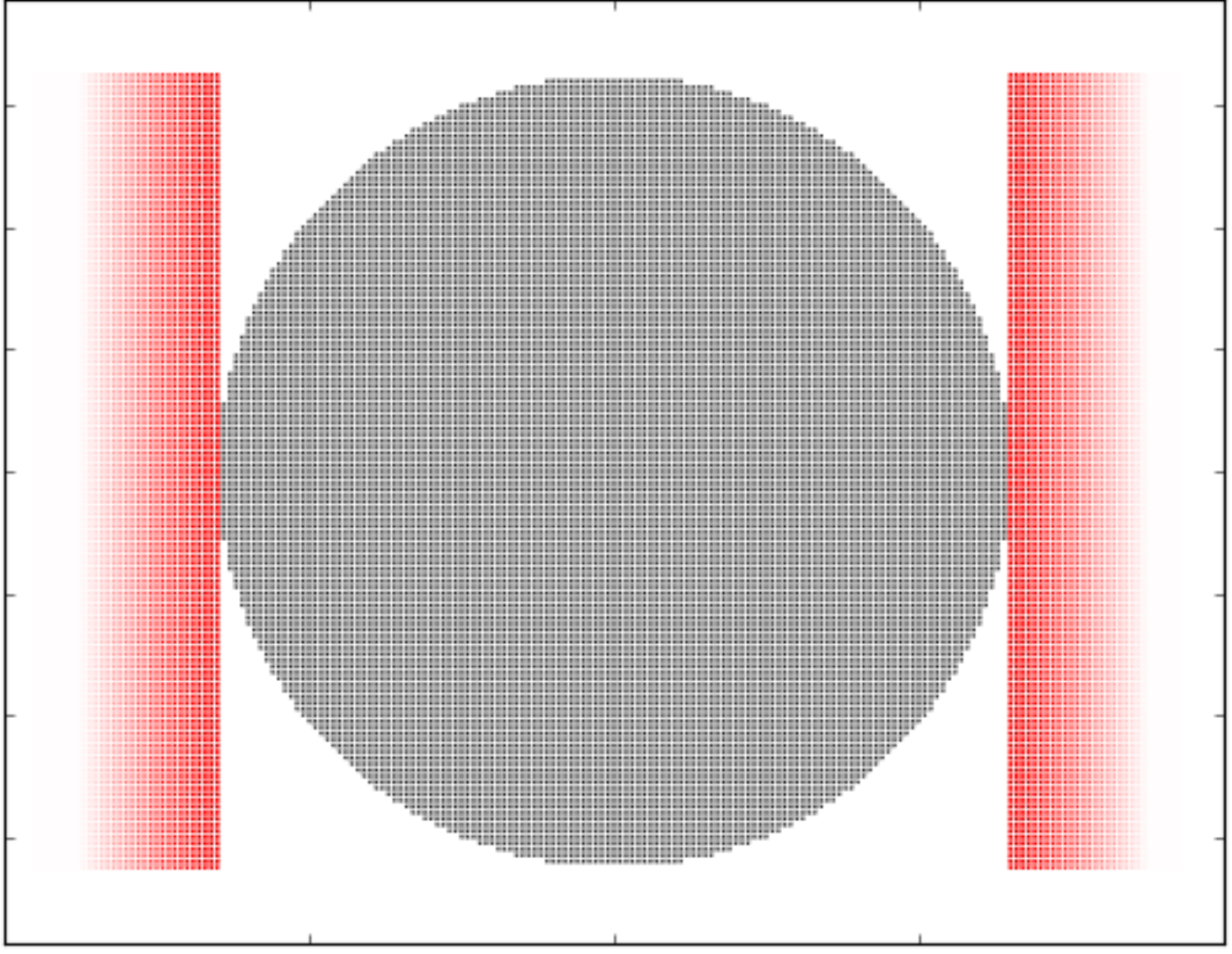}\hspace*{5mm}
    b)\ \includegraphics[height=2.8cm]{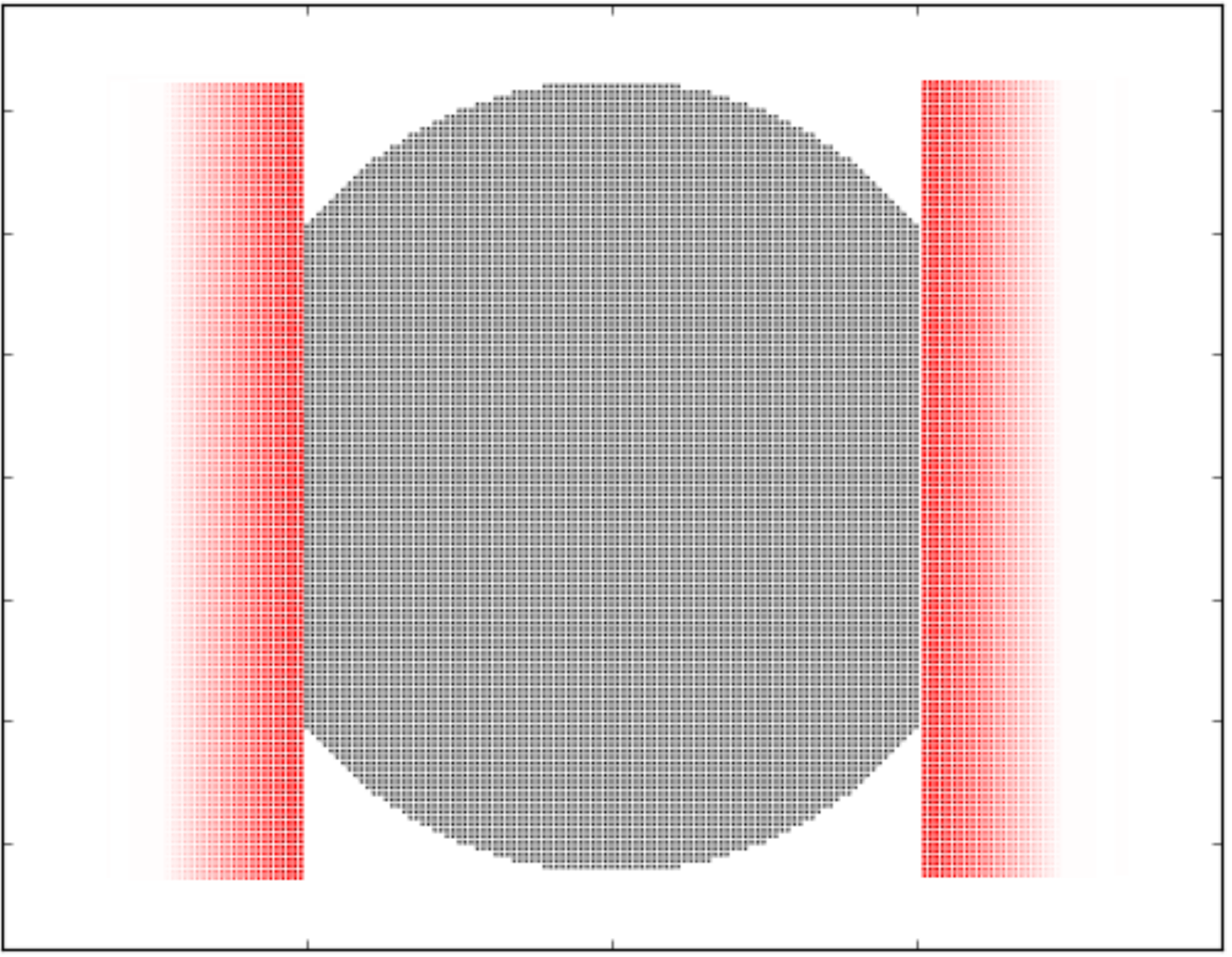}}
  \caption{Geometries of the systems studied with the help of Kwant corresponding to weak (a) and strong (b)
    coupling between the DRN and the leads.\label{kwant-scheme}}
\end{figure}

In approach no. 1 we assume the wave functions in the form given by Eq.~(\ref{eq_psi_nl}). To find the radial
part $R_{nl}(r)$ we apply the Numerov algorithm \cite{numerov} and the shooting method. In approach no. 2 the
two--dimensional Schr\"odinger equation is cast in a matrix form where a discrete basis is used instead of
the continuous real--space position basis. Then, the low--lying states and corresponding energies are
found by diagonalizing the matrix with the help of the Lancz\"os method. In the last approach Kwant program
is used. It is a free software package that uses the Python programming language. It is available at
http://kwant-project.org/. Kwant can be used for numerical calculations related to transport in quantum
systems. One can use it to calculate conductivity, scattering matrix, wave functions, Green's
functions. This package can be used to perform simulation of metals, semiconductors, topological
insulators, superconductors, etc. One can use various lattice types, e.g., square, triangular or
honeycomb lattice. In our approach we define a tight--binding model on a square lattice with only
nearest--neighbor hoppings and the on--site potential given by Eq.~(\ref{potential}). We have infinite
leads attached to both sides of the DRN. By changing the positions of the leads we can 
control the strength of the coupling between the leads and the nanosystem. Fig. \ref{kwant-scheme}
shows the geometry in the cases of weak and strong coupling.
With the possibility of controlling of the DRN--lead coupling we can study how the energy spectrum and the
wave functions are affected by attaching the leads. This is important because the wave functions are needed
to calculate microscopically the tunneling matrix element between states in the DRN ($\Psi_{nl}$) and in leads
($\psi_{\bm k}$)
\begin{equation}
  t_{nl{\bm k}}=\frac{\hbar^2}{2m^*}\int \left[\Psi_{nl}^*({\bm r})\nabla\psi_{\bm k}({\bm r})
    -\psi_{\bm k}^*({\bm r})\nabla\Psi_{nl}({\bm r})\right]\cdot d{\bm S},
  \label{ttt}
\end{equation}
where the integration is performed over a surface separating the DRN and the leads.
Within the Bardeen approach \cite{Bardeen,Tersoff,Tersoff1,Chen,Chen1} it is assumed that 
$\Psi_{nl}$ and $\psi_{k}$ are eigenfunctions
of the DRN and the leads, respectively. However, when the coupling gets stronger the real wave functions are
modified and the Bardeen approach is not valid any more. Below we will show how increasing coupling affects
the wave functions and how it affects conductivity.

\begin{figure*}[h]
  \centerline{a)\hspace*{-3mm}\includegraphics[height=3.3cm]{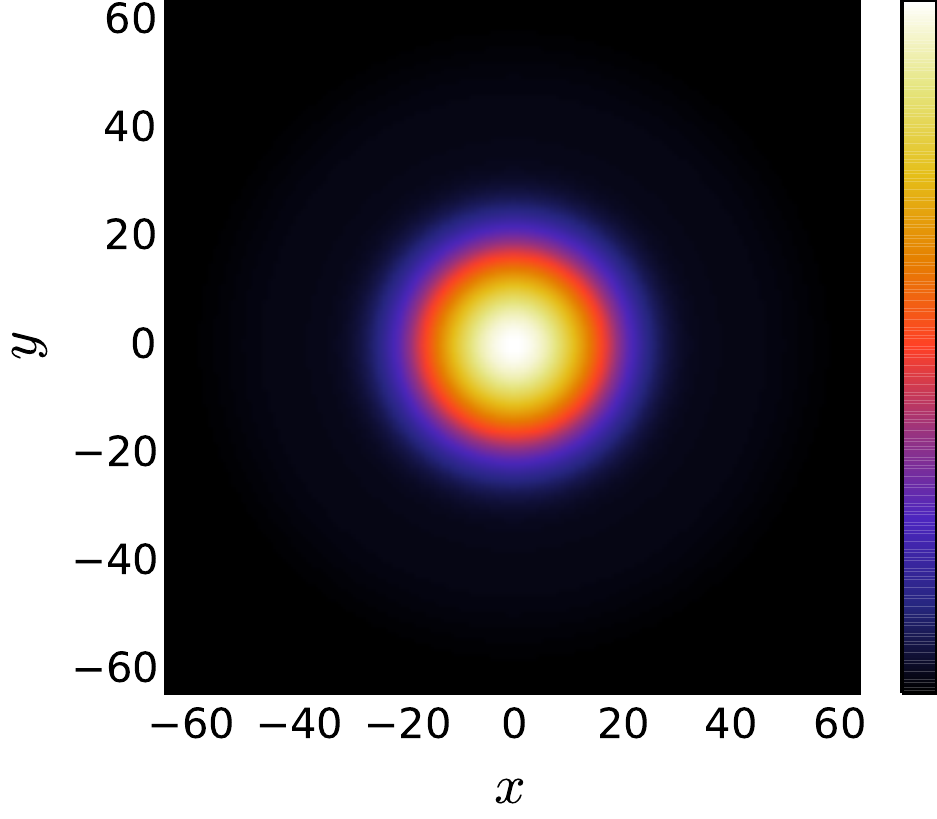}\hspace*{1mm}
    b)\hspace*{-3mm}\includegraphics[height=3.3cm]{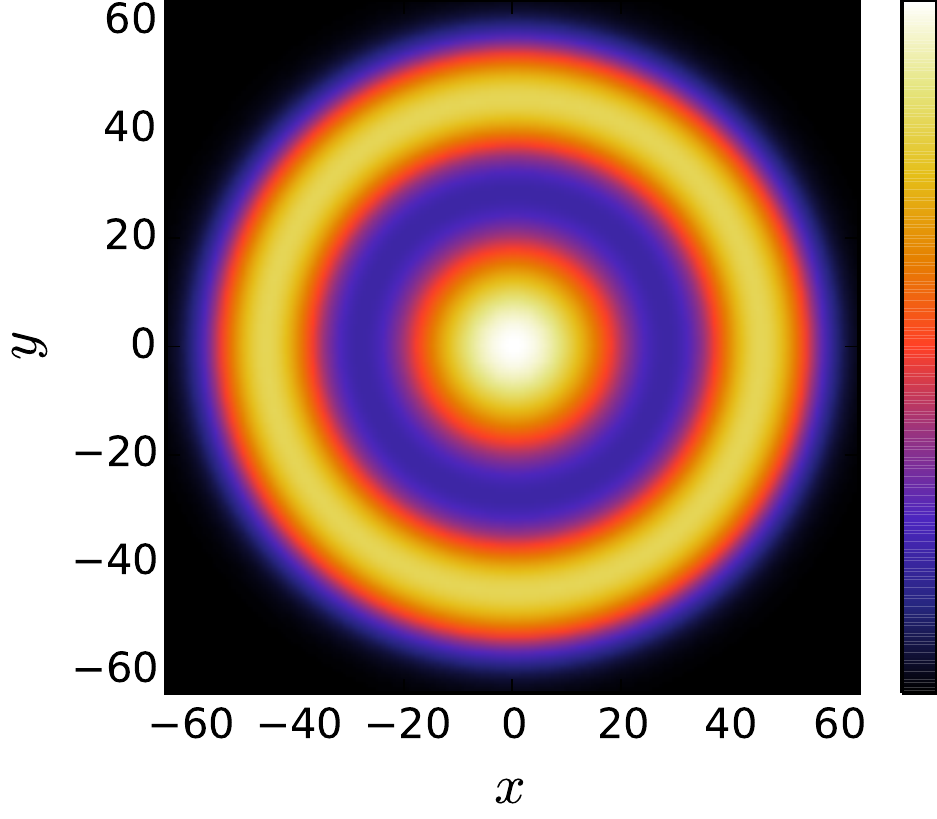}\hspace*{1mm}
    c)\hspace*{-3mm}\includegraphics[height=3.3cm]{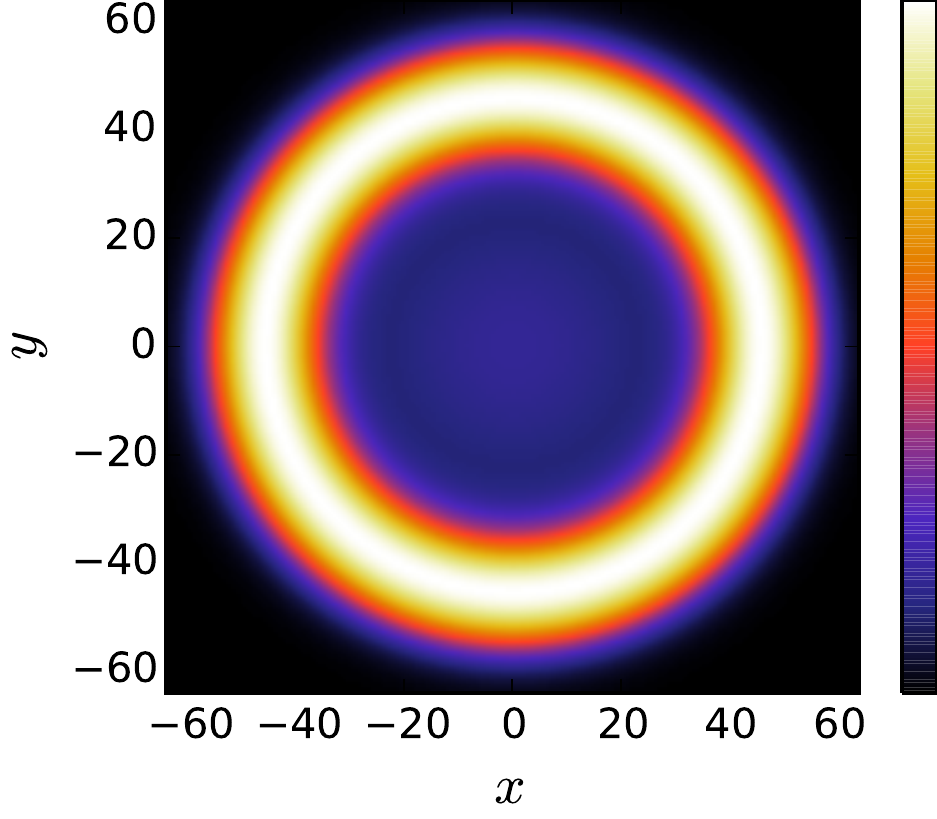}}
  \caption{Absolute value of the ground state $|\Psi_{00}({\bm r})|$ for $V_0=30$ meV and  $V_{\rm QD}$
    equal to 0 (a), 4 meV (b) and 6.5 meV (c).\label{vqd}}
\end{figure*}

But first we want to demonstrate the possibility of wave function engineering in a DRN. By this term we mean
the ability to control the spatial distribution of the electron wave function by changing the shape
of the confining potential. It is possible to control many of the potential parameters but here we will
show the effect of tuning of the position of the QD potential $V_{\rm QD}$. It can be realized by electrical
gating below or above the central part of the DRN. Figure \ref{vqd} shows the distribution of the
ground state wave function for three different values of $V_{\rm QD}$. With increasing position of the
bottom of the QD potential the wave function is moved over from the QD part to the QR part of the DRN.

\begin{figure*}[h]
  \centerline{a)\hspace*{-3mm}\includegraphics[height=3.3cm]{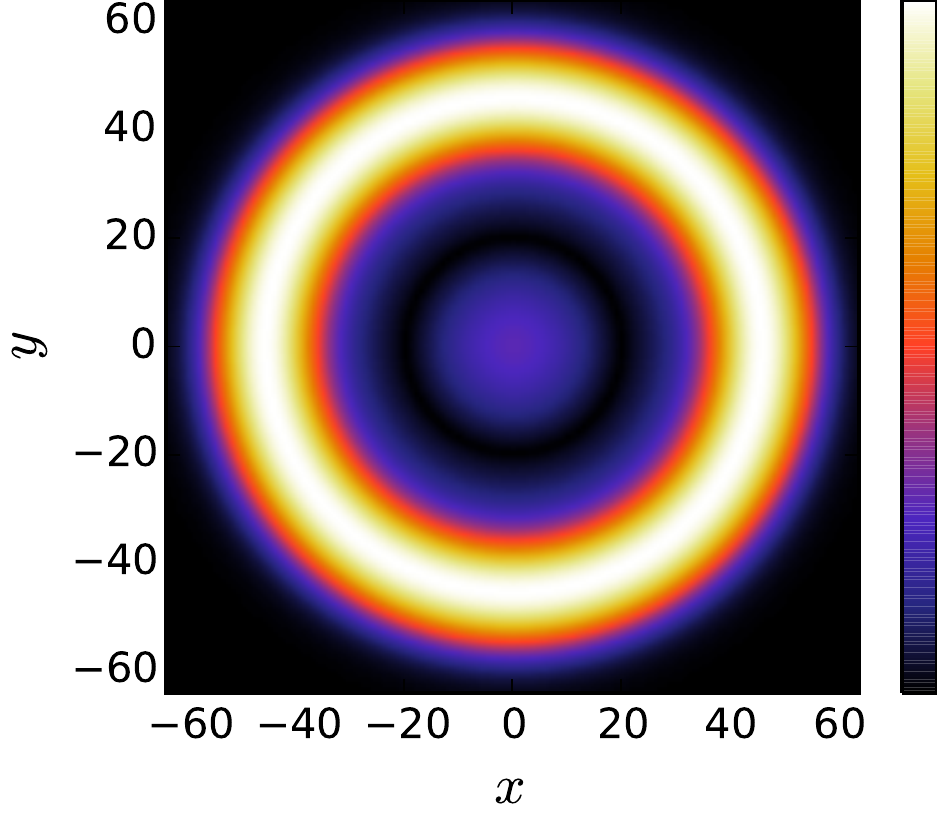}\hspace*{1mm}
    b)\hspace*{-3mm}\includegraphics[height=3.3cm]{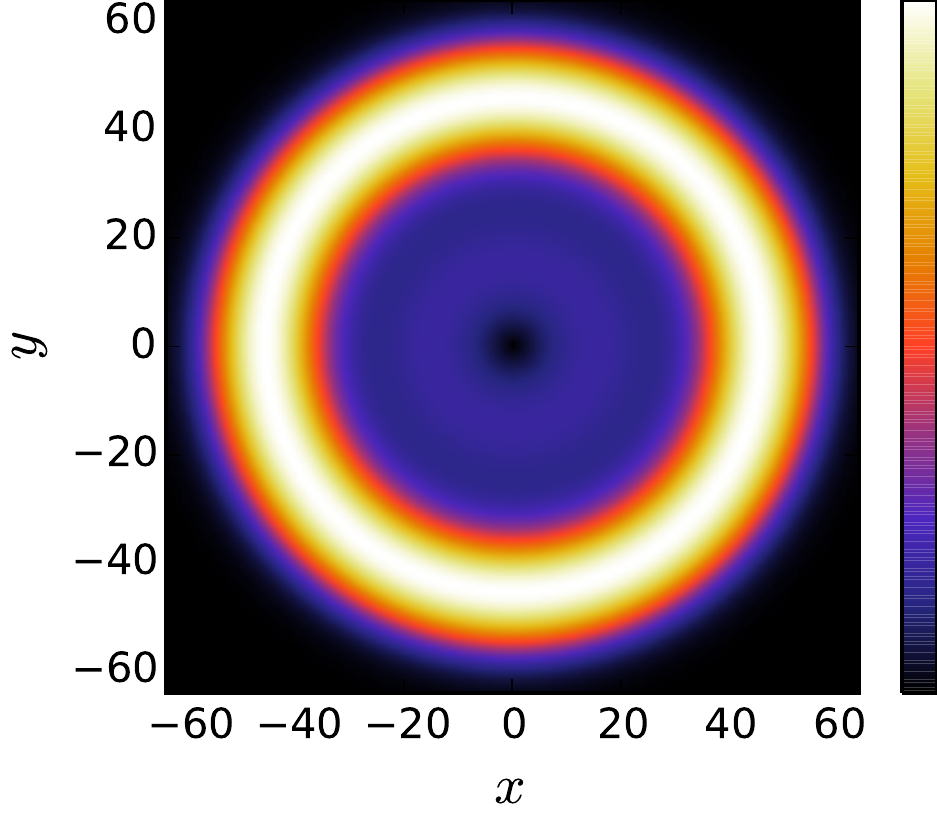}\hspace*{1mm}
    c)\hspace*{-3mm}\includegraphics[height=3.3cm]{fig8}}
  \centerline{\hspace*{3mm}d)\hspace*{-3mm}\includegraphics[height=3.35cm]{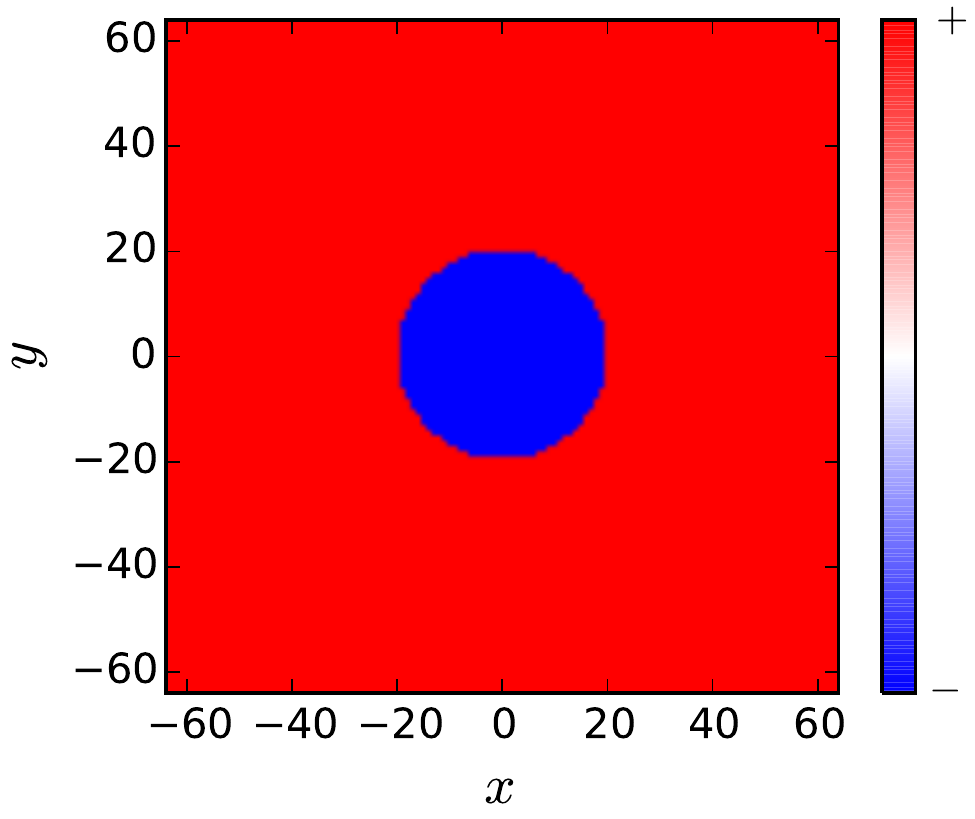}\hspace*{1mm}
    \hspace*{-2mm}e)\hspace*{-3mm}\includegraphics[height=3.35cm]{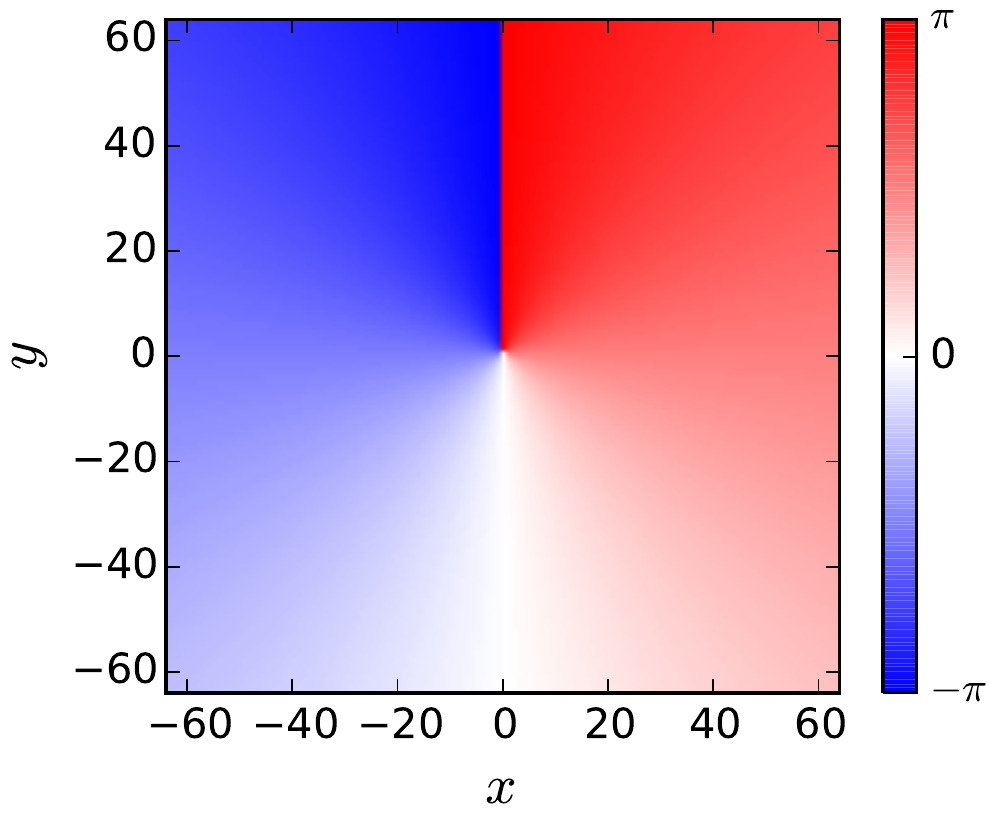}\hspace*{1mm}
    \hspace*{-2mm}f)\hspace*{-3mm}\includegraphics[height=3.35cm]{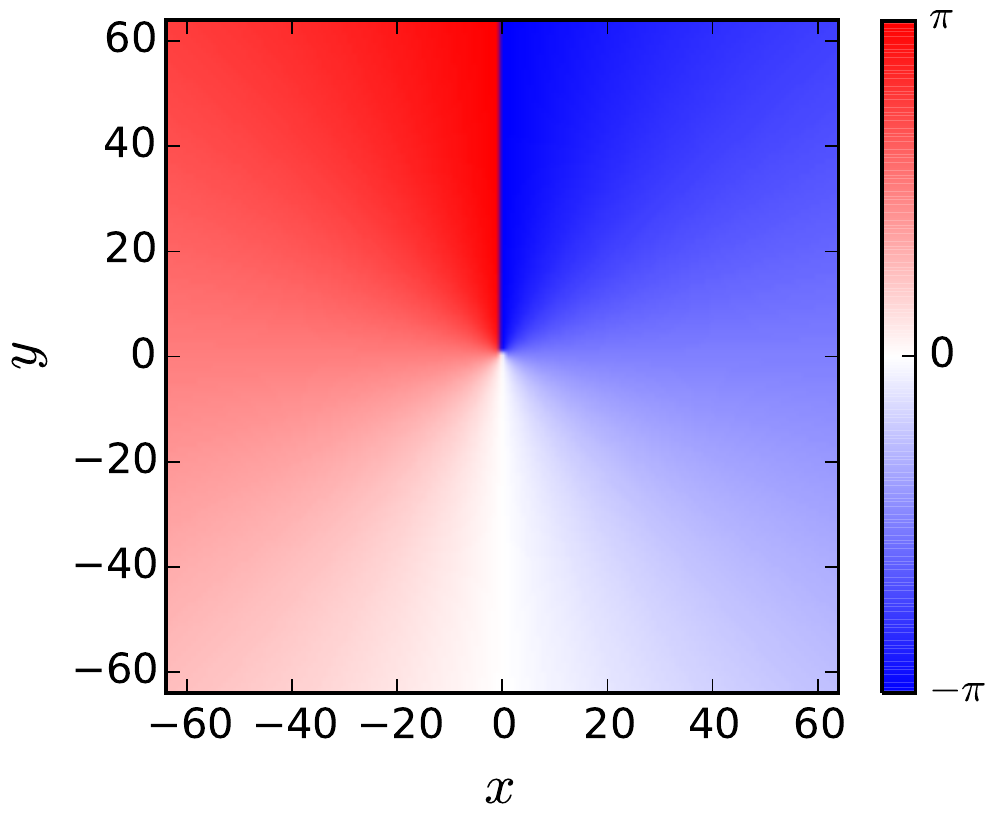}}
  \caption{The absolute value and the phase of the first [(a) and (d)], the second [(b) and (e)] and
    the third [(c) and (f)] excited state of a DRN for $V_0=30$ meV and $V_{\rm QD}=0$. These states
    were calculated within approach no. 1 (see text). The red and blue regions (marked with $+$
    and $-$ on the color bar) in panel (d) represent areas where the (real) wave function presented
    in panel (a) is positive and negative, respectively.\label{ffs1}}
\end{figure*}

The wave functions have been calculated within the approach no. 1, i.e., by solving the Schr\"odinger
equation only for the radial part $R_{nl}(r)$. For all values of $V_{\rm QD}$ the state with $n=0$
and $l=0$ is the ground state. One can see
that for $V_{\rm QD}=0$ the ground state wave function is located in the QD part of the DRN. Then, the
wave function is pushed towards the ring part of the DRN with rising the bottom of the QD potential.
The ground state wave function has a constant phase and its shape is exactly the same in all three
mentioned above approaches. The situation can be different for excited states. Moreover, the spatial
distribution of the wave functions can be significantly affected by even a very weak coupling to leads. 
We start the comparison of the wave functions calculated within different approaches with the case without
leads. Since the ground state in all these methods is almost exactly the same, we start with
the excited states. In Fig. \ref{ffs1} one can see the first ($\Psi_{10}$), the second ($\Psi_{01}$) and
the third ($\Psi_{0,-1}$) excited states (the second and third are degenerated) calculated within approach
no. 1, where the full (complex) wave function is given by Eq. (\ref{eq_psi_nl}). These states were calculated
for $V_0=30$ meV and $V_{\rm QD}=0$. By comparing with Fig. \ref{vqd}a one can infer that these states can be
located in a different part of the DRN than the ground state. As a result the matrix element between
these states and the ground state is small and in Ref. \onlinecite{zipper, MK0} we have demonstrated that
this feature can be used to increase the relaxation time for a spin qubit built on a DRN.

\begin{figure*}[h]
  \centerline{a)\hspace*{-3mm}\includegraphics[height=3.3cm]{fig7}\hspace*{1mm}
    b)\hspace*{-3mm}\includegraphics[height=3.3cm]{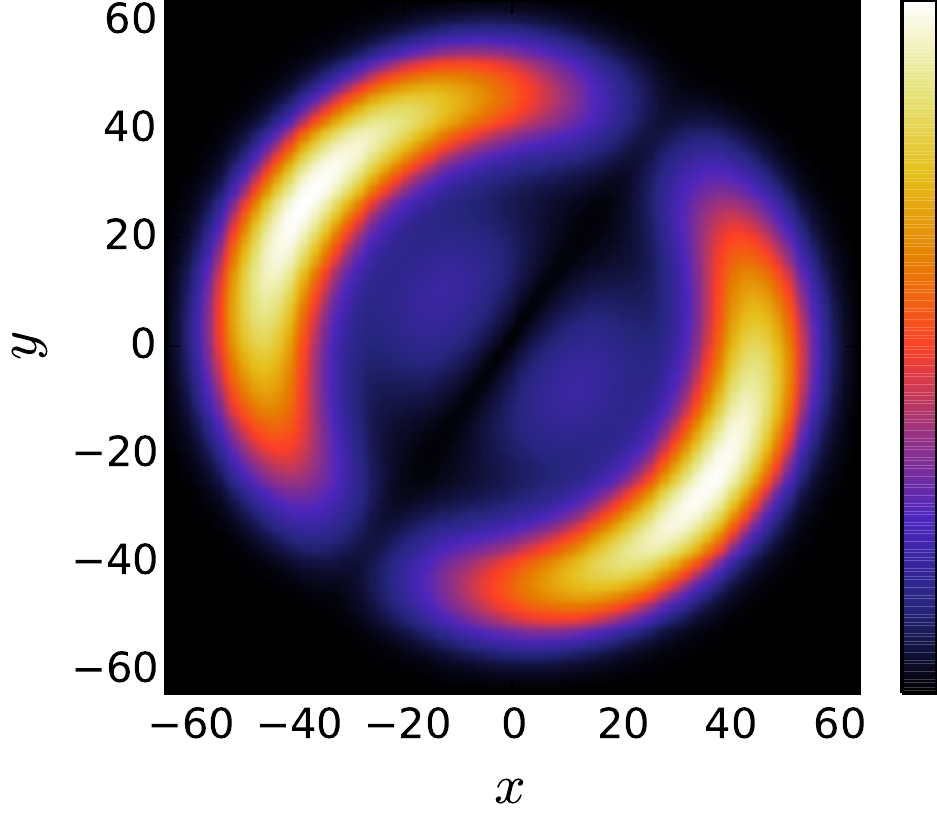}\hspace*{1mm}
    c)\hspace*{-3mm}\includegraphics[height=3.3cm]{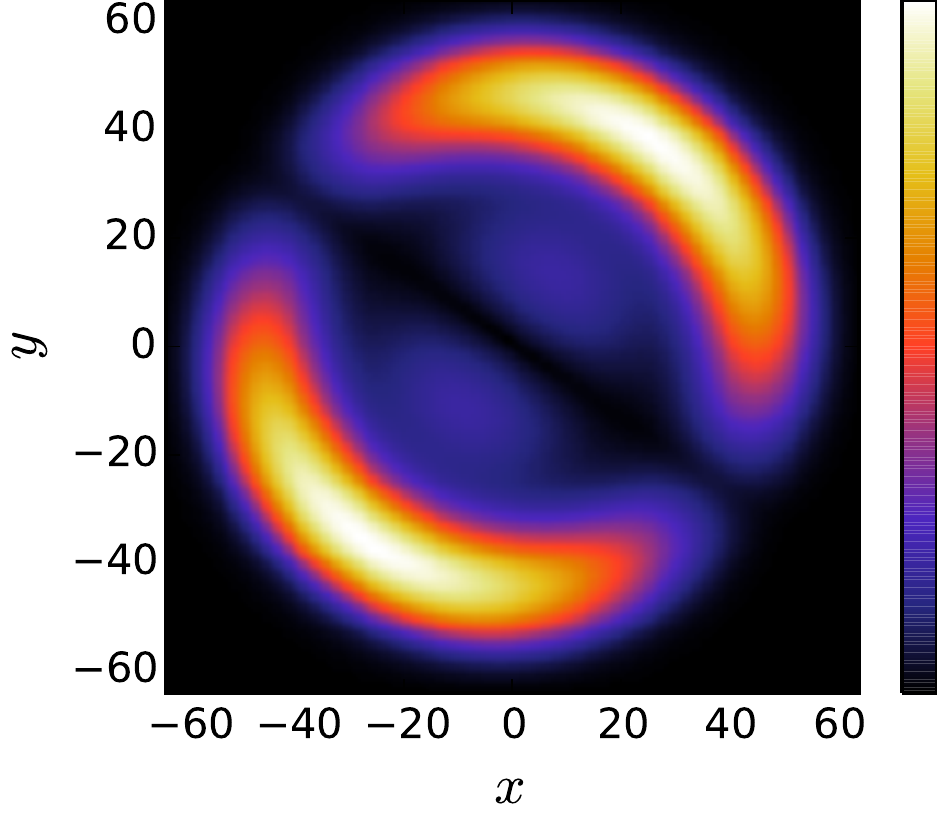}}
  \centerline{\hspace*{2mm}d)\hspace*{-3mm}\includegraphics[height=3.37cm]{fig11}\hspace*{1mm}
    \hspace*{-2mm}e)\hspace*{-3mm}\includegraphics[height=3.37cm]{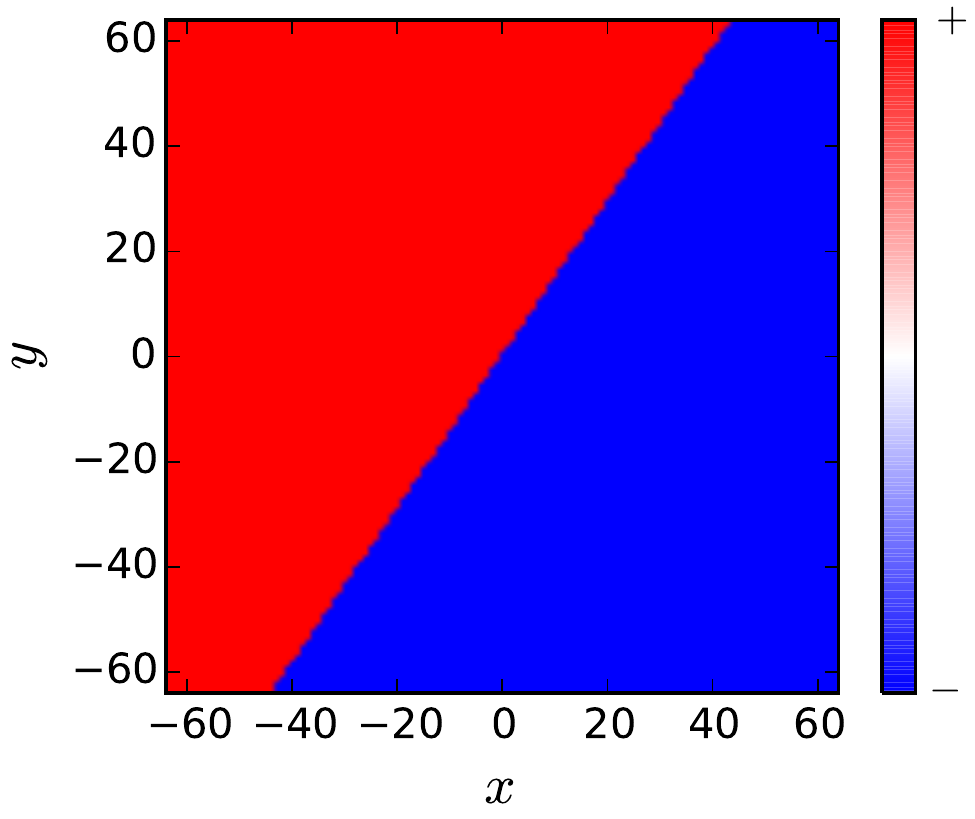}\hspace*{1mm}
    \hspace*{-2mm}f)\hspace*{-3mm}\includegraphics[height=3.37cm]{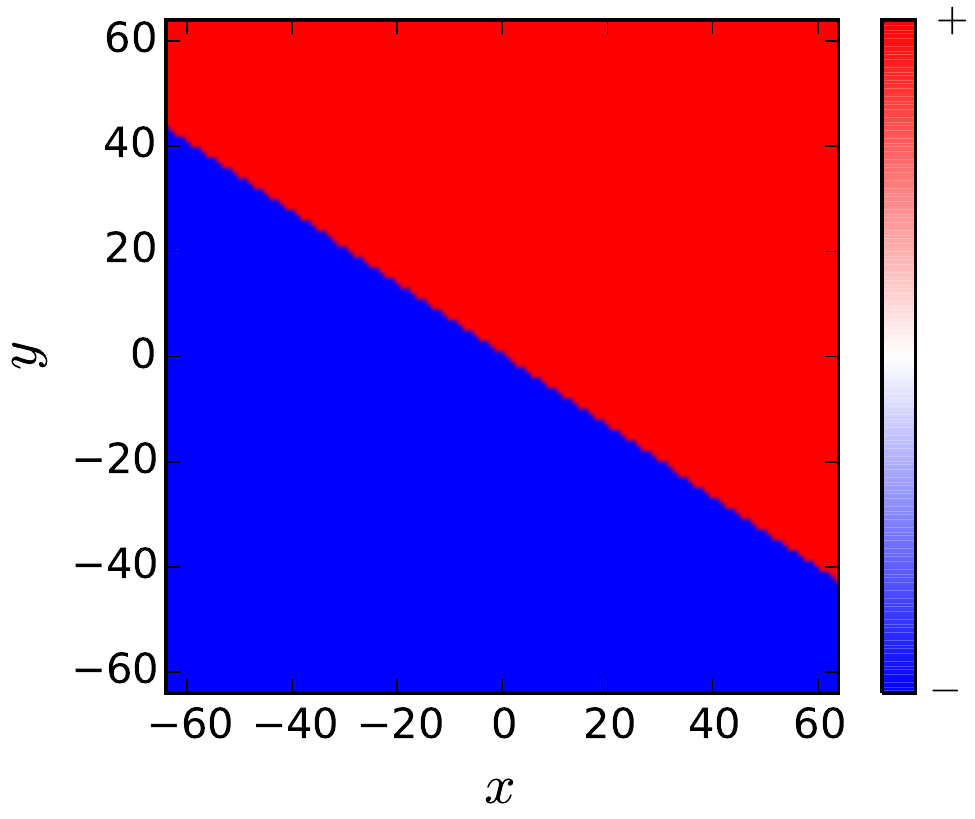}}
  \caption{The same as in Fig. \ref{ffs1}, but calculated within approach no. 2.
    \label{ffs2}}
\end{figure*}

\begin{figure*}[htb]
  \centerline{a)\hspace*{-3mm}\includegraphics[height=3.3cm]{fig7}\hspace*{1mm}
    b)\hspace*{-3mm}\includegraphics[height=3.3cm]{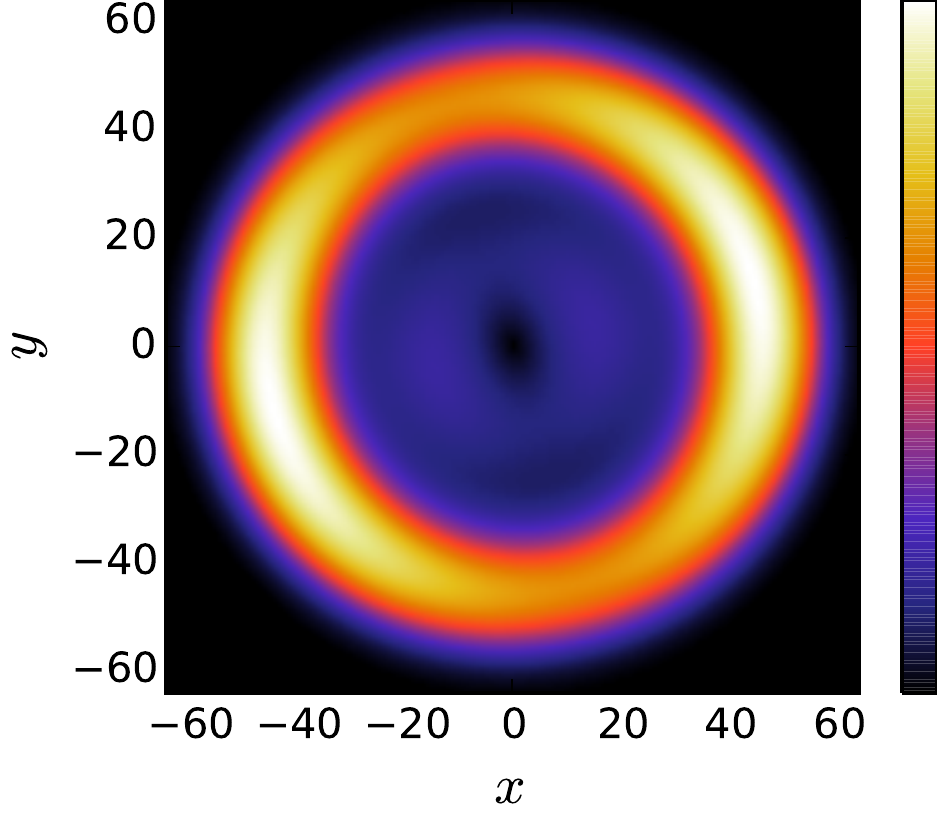}\hspace*{1mm}
    c)\hspace*{-3mm}\includegraphics[height=3.3cm]{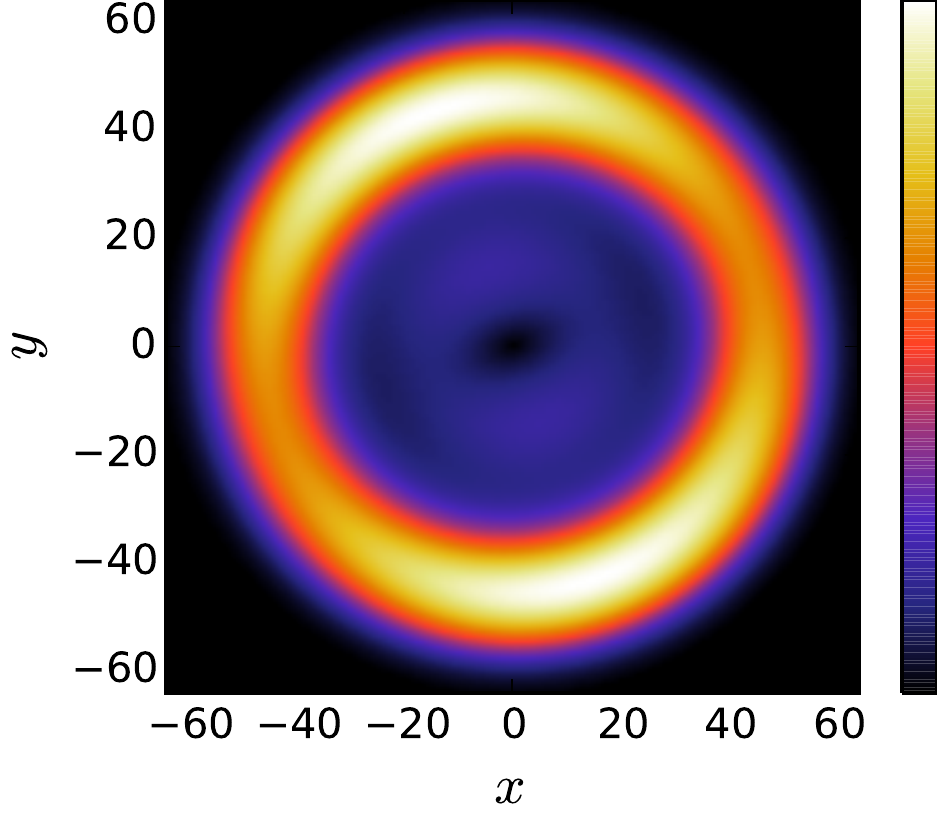}} 
  \centerline{\hspace*{3mm}d)\hspace*{-3mm}\includegraphics[height=3.33cm]{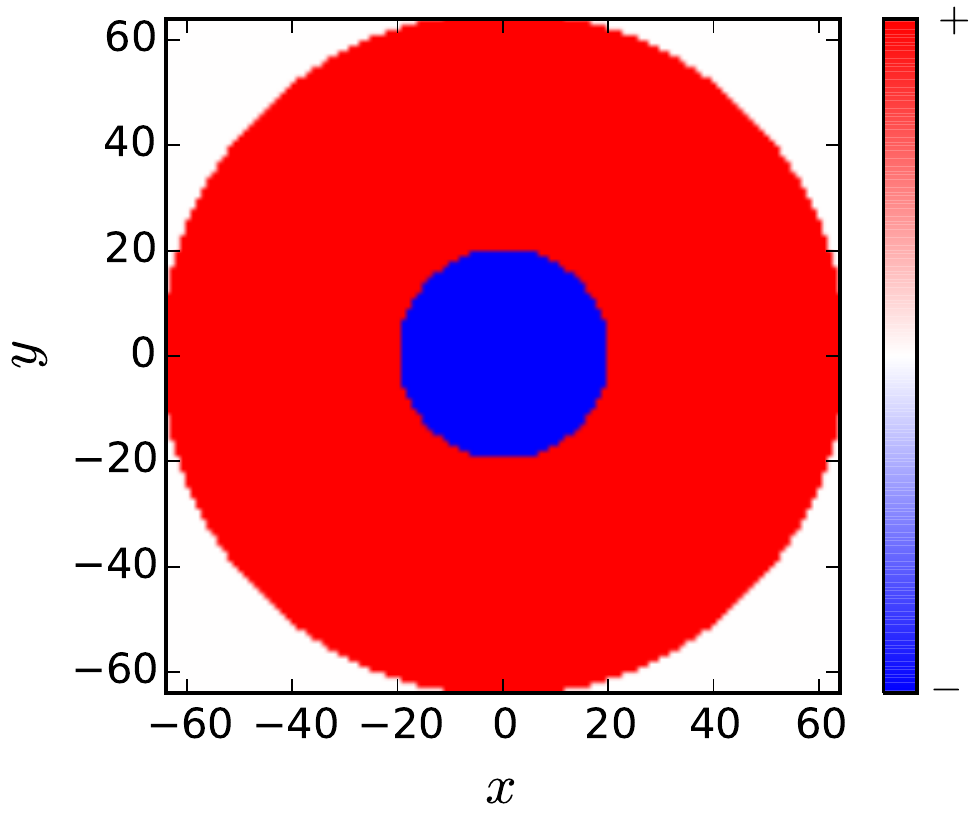}\hspace*{1mm}
    \hspace*{-2mm}e)\hspace*{-3mm}\includegraphics[height=3.33cm]{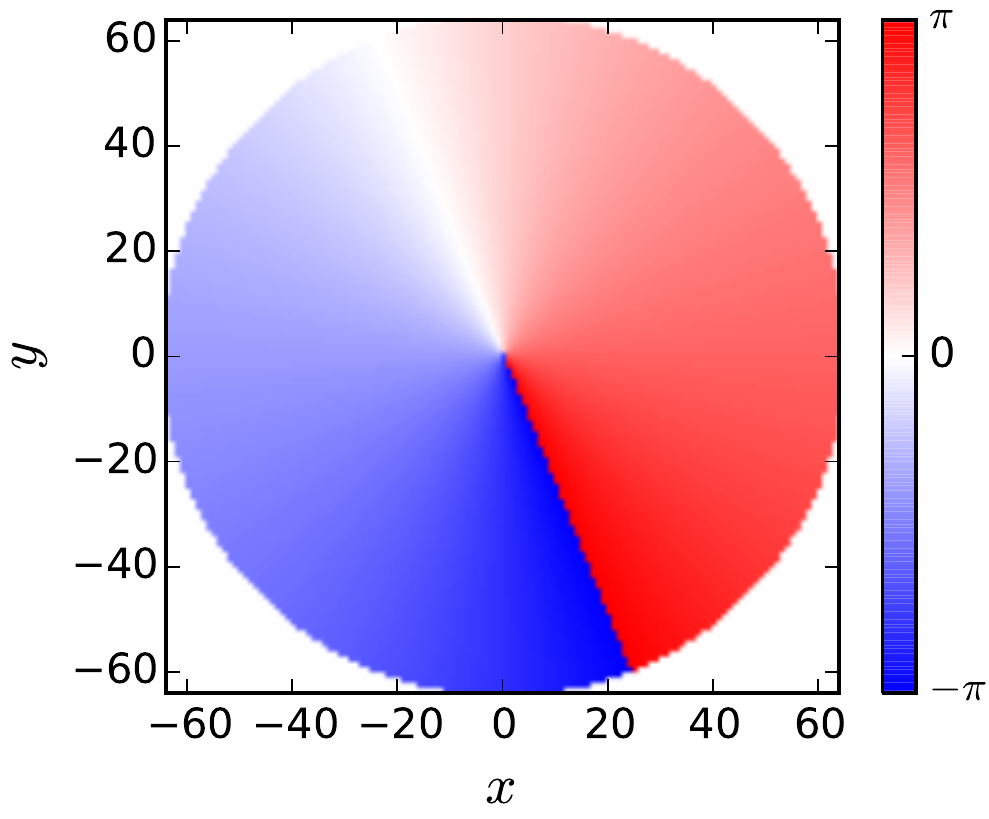}\hspace*{1mm}
    \hspace*{-2mm}f)\hspace*{-3mm}\includegraphics[height=3.33cm]{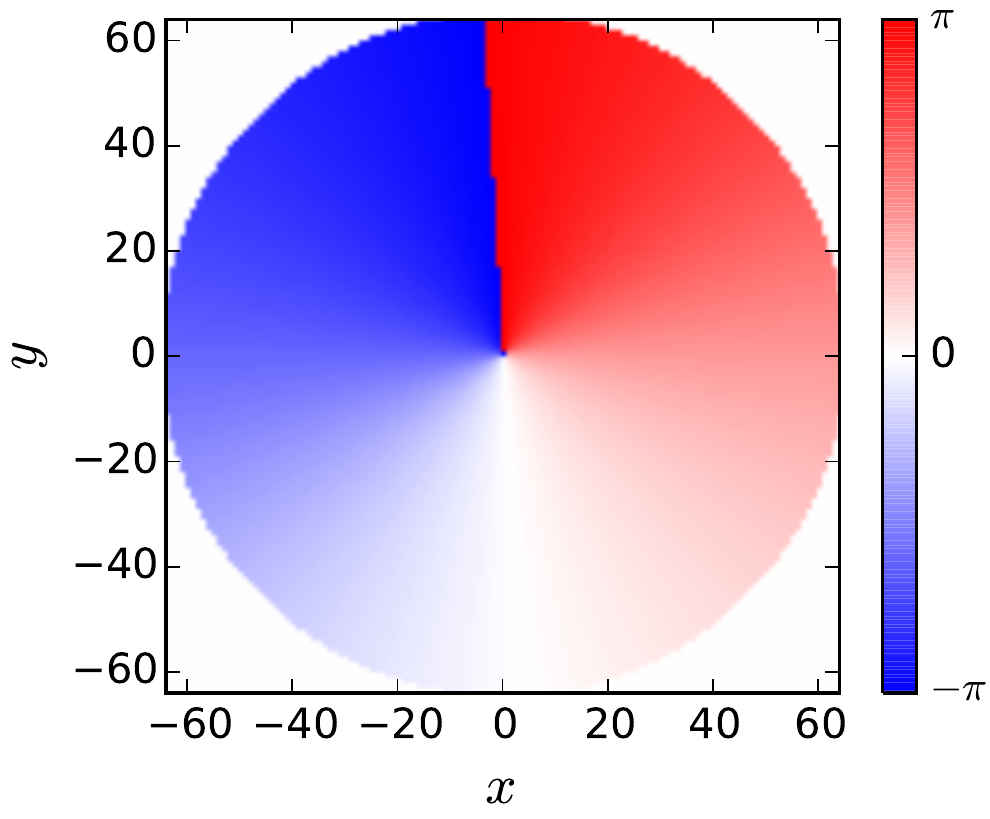}}
  \caption{The same as in Fig. \ref{ffs1}, but calculated within approach no. 3.\label{ffs3}}
\end{figure*}

\begin{figure}[htb]
  \centerline{a)\hspace*{-2mm}\includegraphics[height=3.3cm]{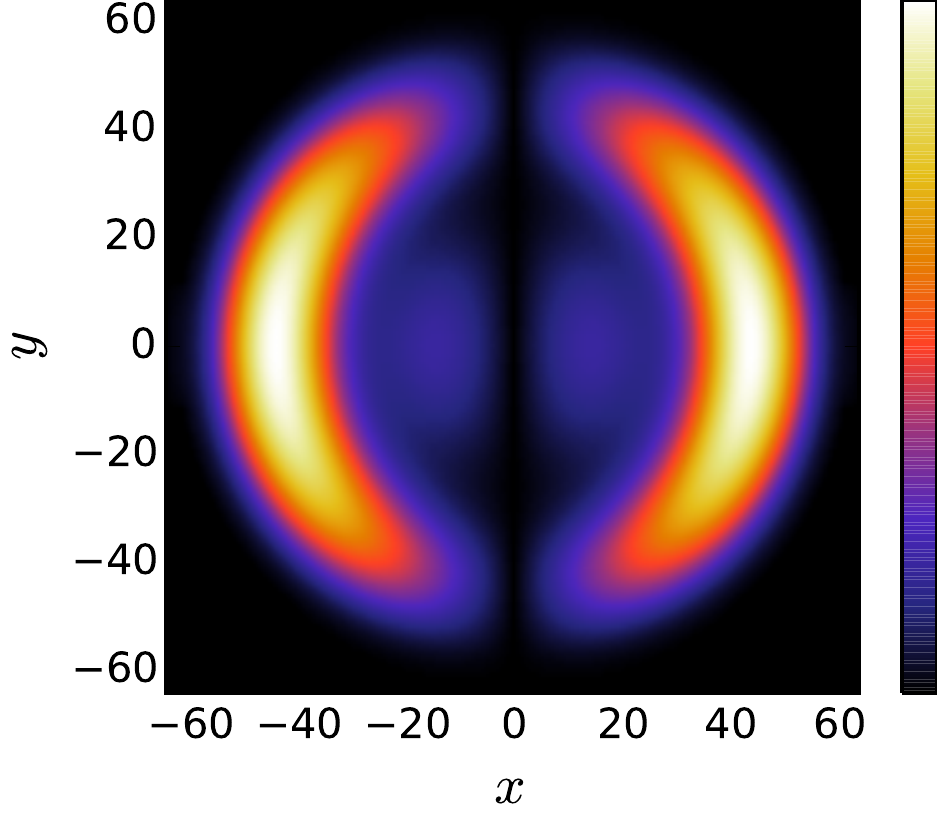}\hspace*{1mm}
    b)\hspace*{-2mm}\includegraphics[height=3.3cm]{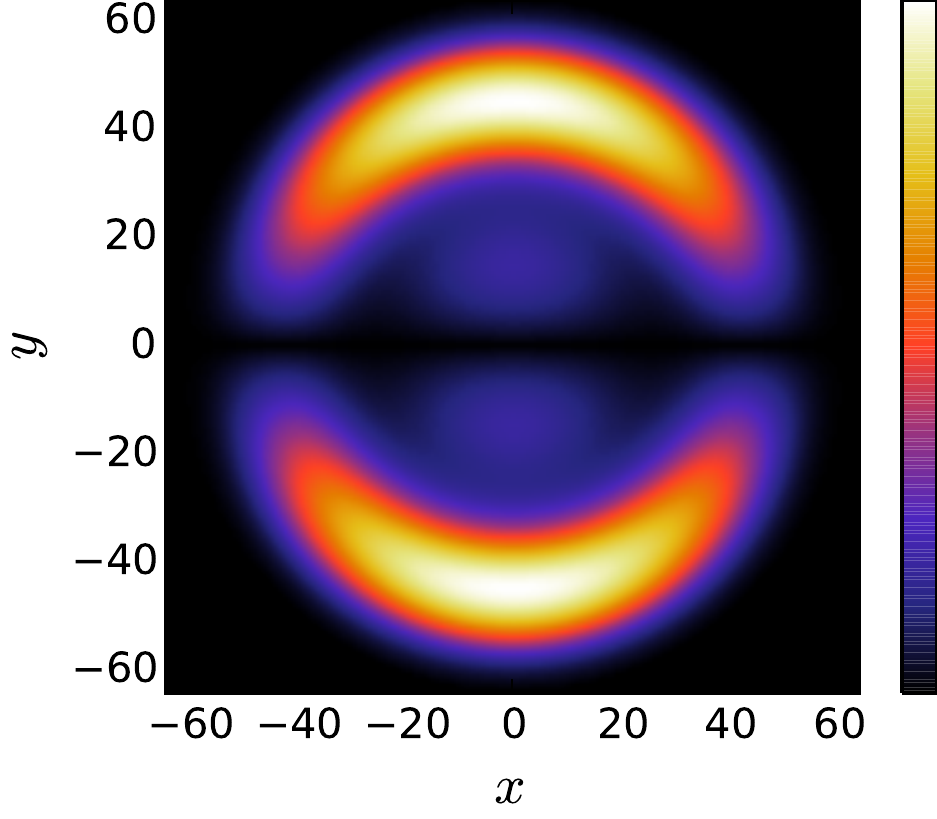}}
  \centerline{\hspace*{3.2mm}c)\hspace*{-2mm}\includegraphics[height=3.35cm]{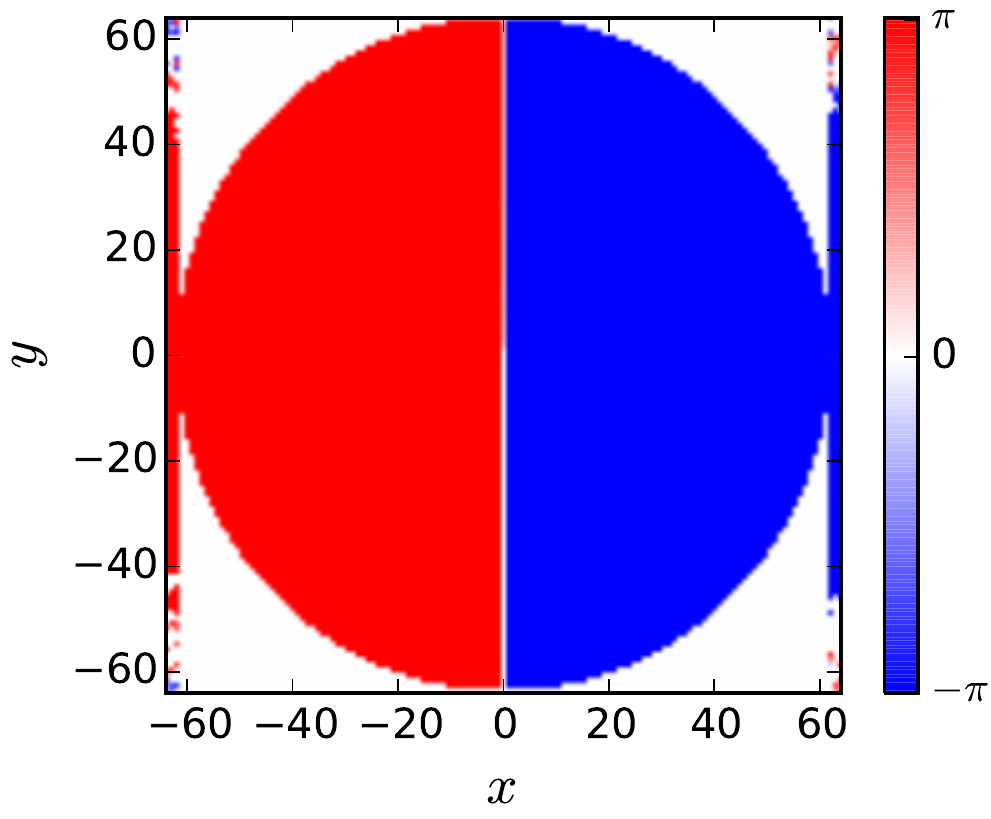}\hspace*{-1.5mm}
    d)\hspace*{-2mm}\includegraphics[height=3.35cm]{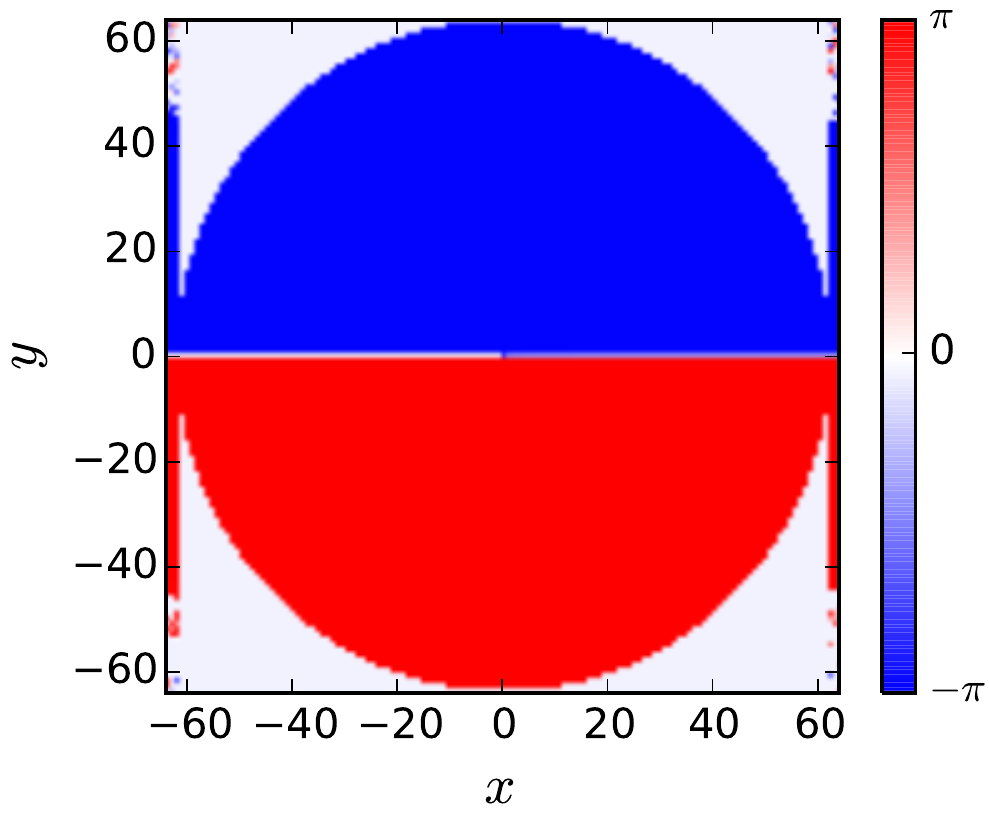}}
  \caption{The same as in Fig. \ref{ffs3}b,e,c,f, but for a system very weakly coupled to leads.
    \label{scon}}
\end{figure}

Fig. \ref{ffs2} shows the same states calculated by solving the Schr\"odinger equation discretized
in two dimensions and then diagonalized with the help of the Lancz\"os method (approach no. 2). In this
case the wave functions are real.
In this approach the cylindrical symmetry is broken by the discretization of the Schr\"odinger equation.
Therefore $n$ and $l$ are not good quantum numbers any more. Of course, the second and the third excited
states are still degenerated. One can check that these wave functions can be expressed as a linear
combination of $\Psi_{0,1}$ and $\Psi_{0,-1}$ (up to a phase factor, see Figs. \ref{ffs1}b,e and \ref{ffs1}c,f):
\begin{eqnarray}
  \Psi_{\rm Fig. \ref{ffs2}b,e}&=&\frac{1}{\sqrt{2}}\left(\Psi_{0,1}+\Psi_{0,-1}\right),\\
  \Psi_{\rm Fig. \ref{ffs2}g,f}&=&\frac{1}{\sqrt{2}i}\left(\Psi_{0,1}-\Psi_{0,-1}\right).
\end{eqnarray}
Excited states calculated from the tight--binding model (approach no. 3) are resented in Fig. \ref{ffs3}.
The wave functions are complex (like in approach no. 1), but the cylindrical symmetry of the Hamiltonian
is broken (like in approach no. 2). The first excited state, which possesses (at least approximately) the
cylindrical symmetry is the same in all the approaches. However, in the next two states the amplitude of the
wave functions is angle--dependent, but the modulation is much weaker than in approach no. 2. Namely, the
wave functions do not have nodes in any direction and therefore their phase cannot change discontinuously
like in approach no. 2. What is interesting, this holds true only for an insulated system. If the system
is even very weakly coupled to leads, the wave functions become real with nodes, like in approach no. 2.
Such situation is presented in Fig. \ref{scon}.

\begin{figure*}[htb]
  \centerline{a)\hspace*{-3mm}\includegraphics[height=3.35cm]{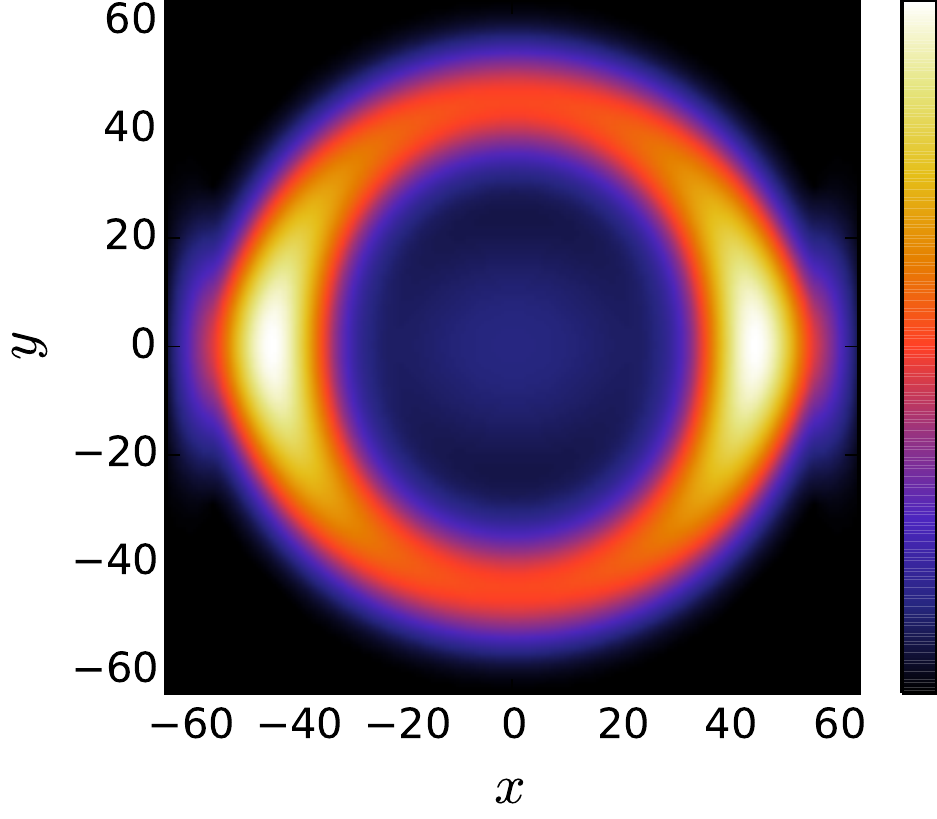}
    b)\hspace*{-3mm}\includegraphics[height=3.35cm]{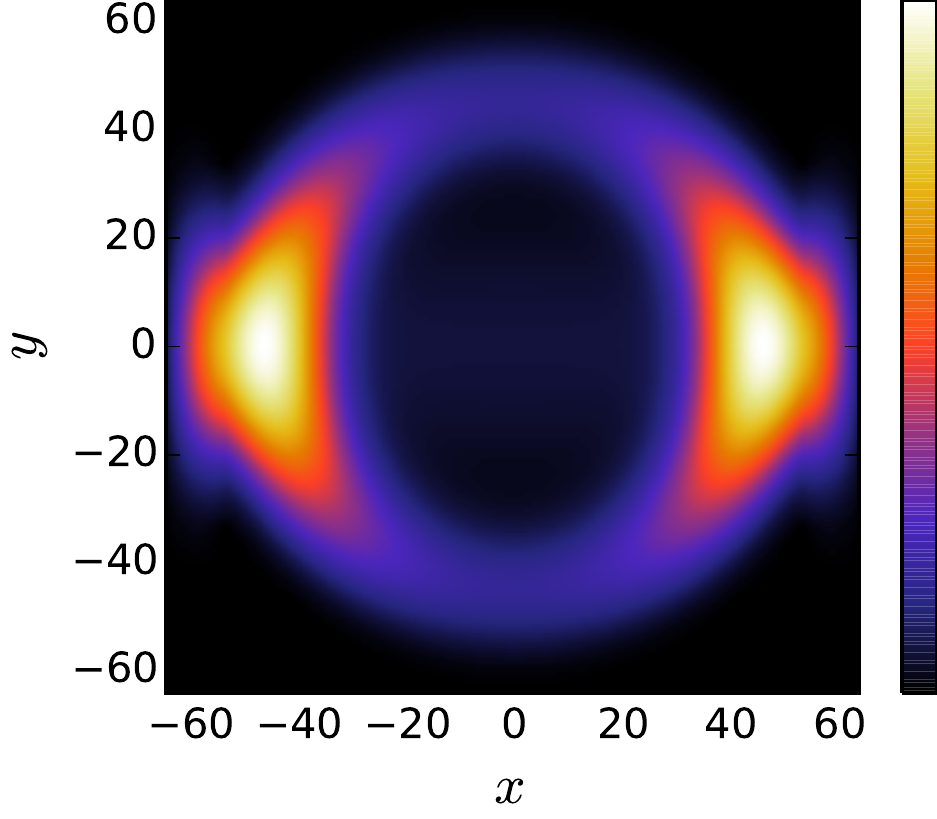}
    c)\hspace*{-3mm}\includegraphics[height=3.35cm]{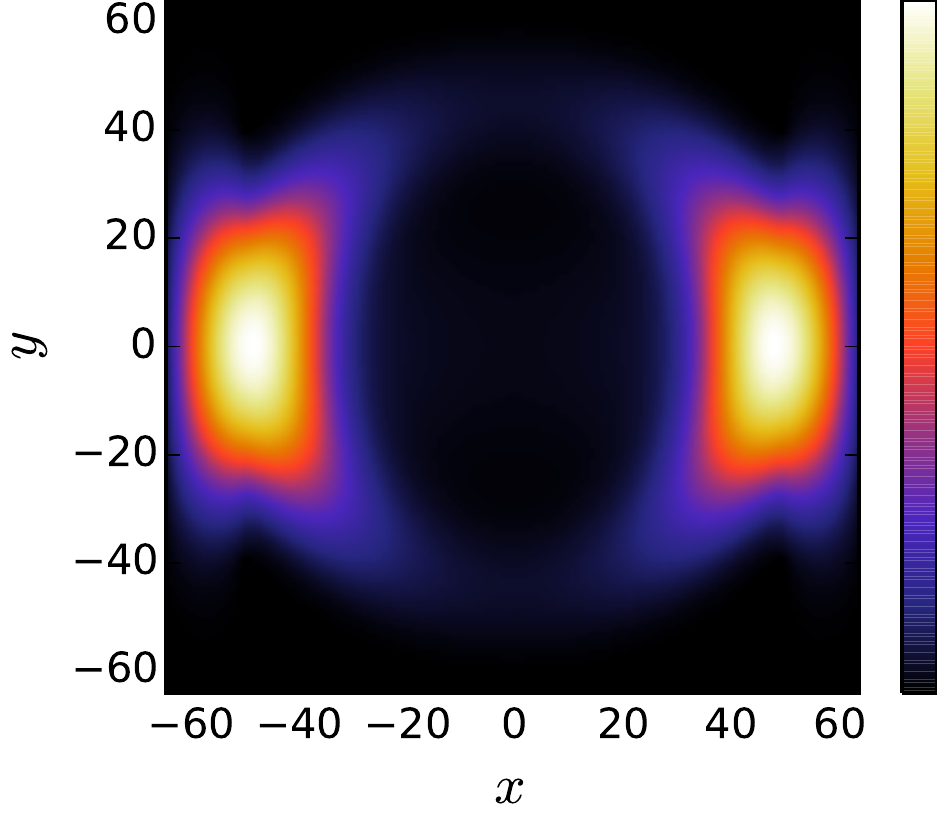}}
  \caption{The ground state wave function for $V_{\rm QD}=6$ meV in the case of weak (a), average (b),
    and strong (c) coupling between the DRN and the leads.\label{coupling}}
\end{figure*}

If the coupling is very weak, it only breaks the underlying cylindrical symmetry of the original Hamiltonian
but the character of the states remains unaffected. The degeneracy of the second and the third excited
states shown in Fig.~\ref{scon} corresponds to the degeneracy of states $\Psi_{01}$ and $\Psi_{0,-1}$.
However, when the coupling increases the change of the geometry becomes important and the spatial
distribution of the wave functions changes as well. The changes are pronounced mainly for states which
are located in the ring part of the DRN, i.e., close to the areas where the leads are attached.
Fig. \ref{coupling} demonstrates how the ground state for $V_{\rm QD}=6$ meV (which even in the
absence of coupling to leads is located entirely in the QR) evolves with increasing coupling.

One can see there that in the case of strong coupling the wave functions are completely different from
these in an insulated system and therefore in this regime the Bardeen approach cannot be valid. In order
to determine the regime where this approximation can be used, in the next section we compare
the coupling constants and conductivities calculated within the Bardeen approach and with the help
of the Kwant package for a tight--binding model.

\section{The dot--ring nanostructure with leads}

In this section we study a system composed of the DRN and weakly coupled source and drain leads
with $\Gamma_S$ and $\Gamma_D$ as the subsequent tunneling rates. We focus here on the case of equal tunnel
barriers, i.e., $ \Gamma_S =\Gamma_D = \Gamma$. Within the first method we calculate the tunnel rate for
tunneling through state $\Psi_{nl}({\bm r})$ as
\begin{equation}
  \Gamma_{nl}(\varepsilon)=2\pi\sum_{\bm k}|t_{nl{\bm k}}|^2\delta\left(\varepsilon-\varepsilon_{\bm k}\right),
  \label{gamma_eq}
\end{equation}
where $t_{nl{\bm k}}$ is given  by Eq. (\ref{ttt}). In the following we restrict our study only to tunneling
through the ground state ($n=0,\:l=0$) and skip the indices ``00'' in $\Gamma_{00}$. Note that the situation
where only the ground state is in the bias window $\mu_S-\mu_D$ not always can be realized because for some
values of $V_{\rm QD}$ energy levels cross.

\begin{figure}[h]
  \centerline{\includegraphics[width=0.5\textwidth]{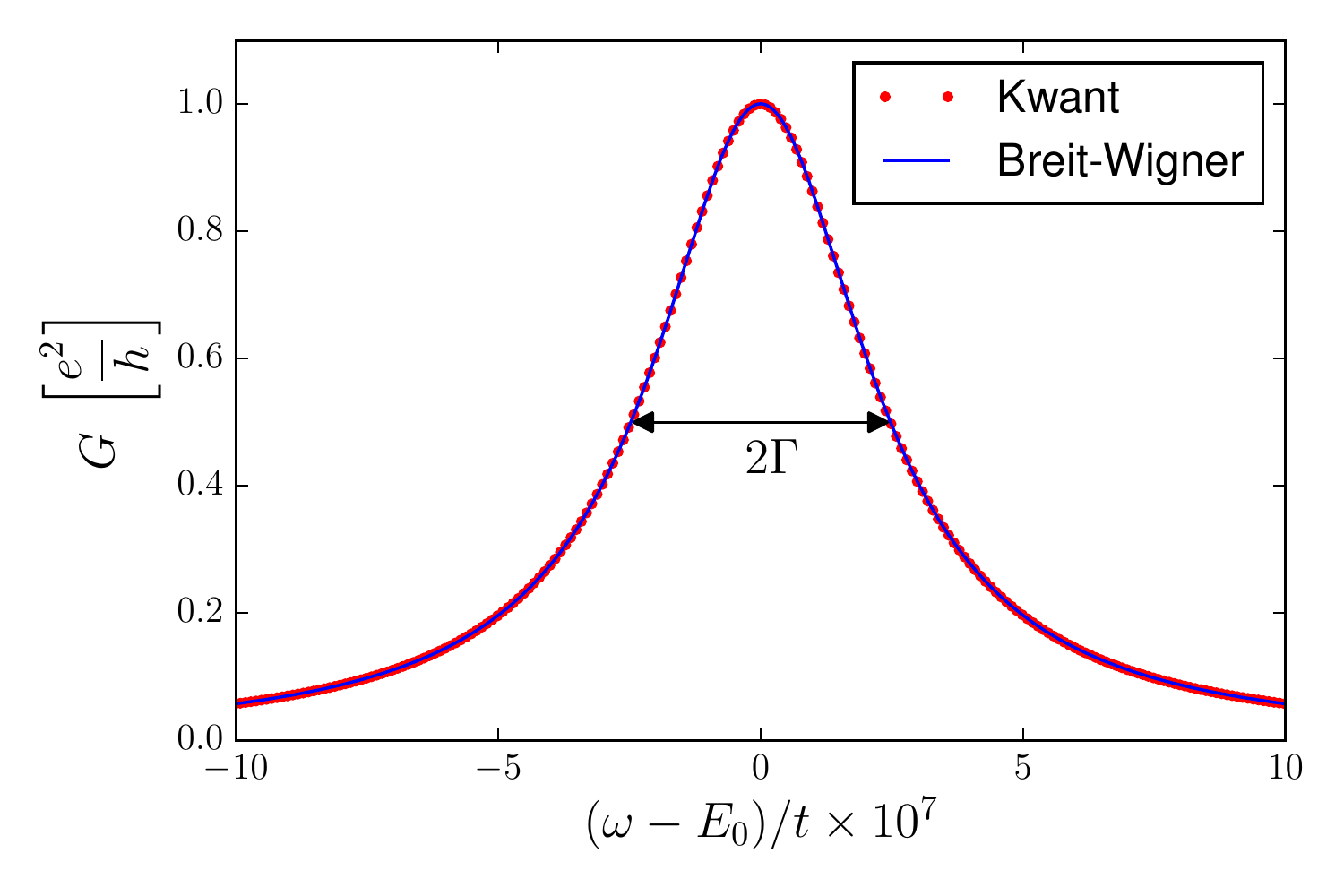}}
  \caption{Fit of the Breit-Wigner formula to a conductance peak for the ground state calculated
    with the help of Kwant for $V_{\rm QD}=-1$ meV. In this case the ground state is located in the QD
    part of the DRN and therefore is very weakly coupled to the leads. Additionally, the leads are
    relatively far from the center of the DRN in the position presented in Fig. \ref{kwant-scheme}a.
    \label{B-W}}
\end{figure}

Since package Kwant does not allow one to calculate the tunneling rates $\Gamma$ directly, we use the
Breit-Wigner formula \cite{BW,BW1,BW2} for the energy-dependent transmission that gives the
half--width--at--half--maximum of the conductance peak equal to $\Gamma$:\footnote{Our studies are
  restricted to low temperatures  $kT\ll \Gamma$ so the thermal broadening of the conductance peak can
  be neglected.}
\begin{equation}
  G(\omega)=\frac{e^2}{h}\frac{\Gamma^2}{(\omega-E_0)^2+\Gamma^2}
\end{equation}
This fit, however, works only for separate peaks in a system with relatively weakly coupled leads.
Fig. \ref{B-W} shows how well the Breit-Wigner formula describes the conductance peak
in this regime.
 For a stronger coupling, when the peak broadening is so large that adjoining peaks
overlap with each other, this formula does not give good results.

\begin{figure}[htb]
  \centerline{\includegraphics[width=0.5\textwidth]{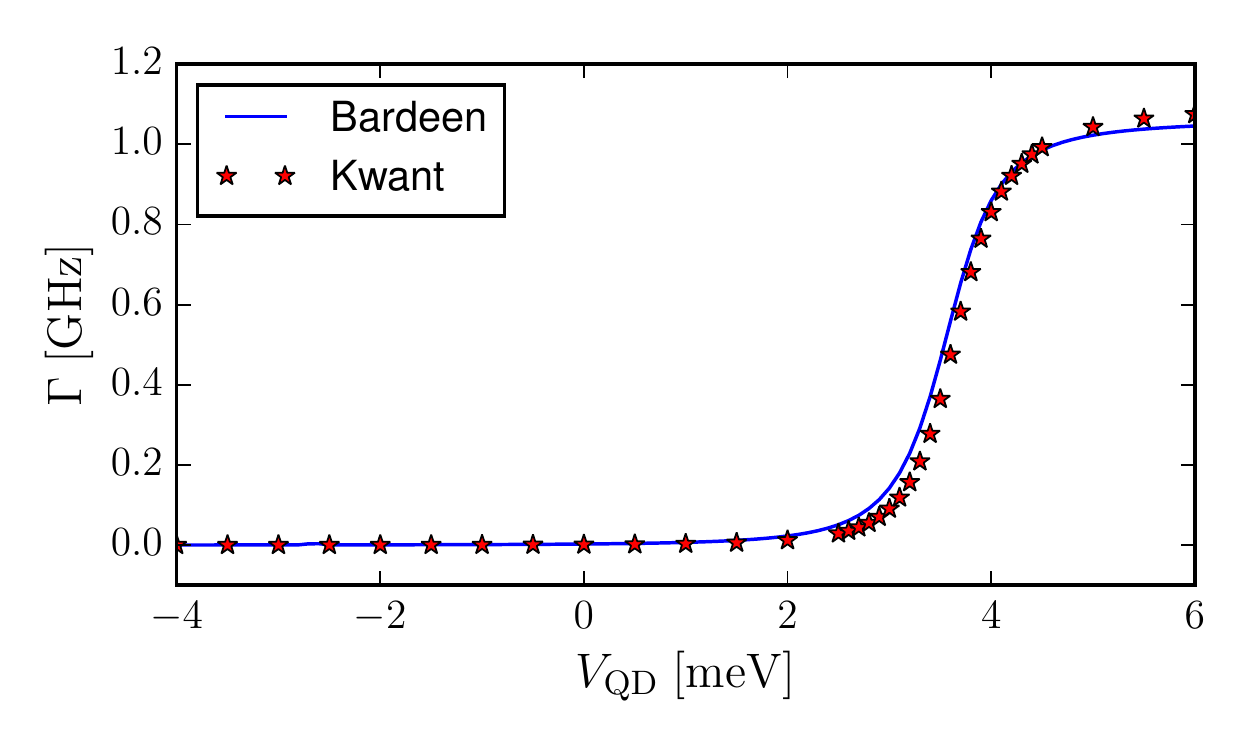}}
  \caption{Comparison of the tunneling rate between the lead and the ground state as a function
    of $V_{\rm QD}$. The solid line and the stars represent results of the Bardeen approach and
    Kwant, respectively.\label{fig36}}
\end{figure}

\begin{figure}[htb]
  \centerline{\includegraphics[width=0.5\textwidth]{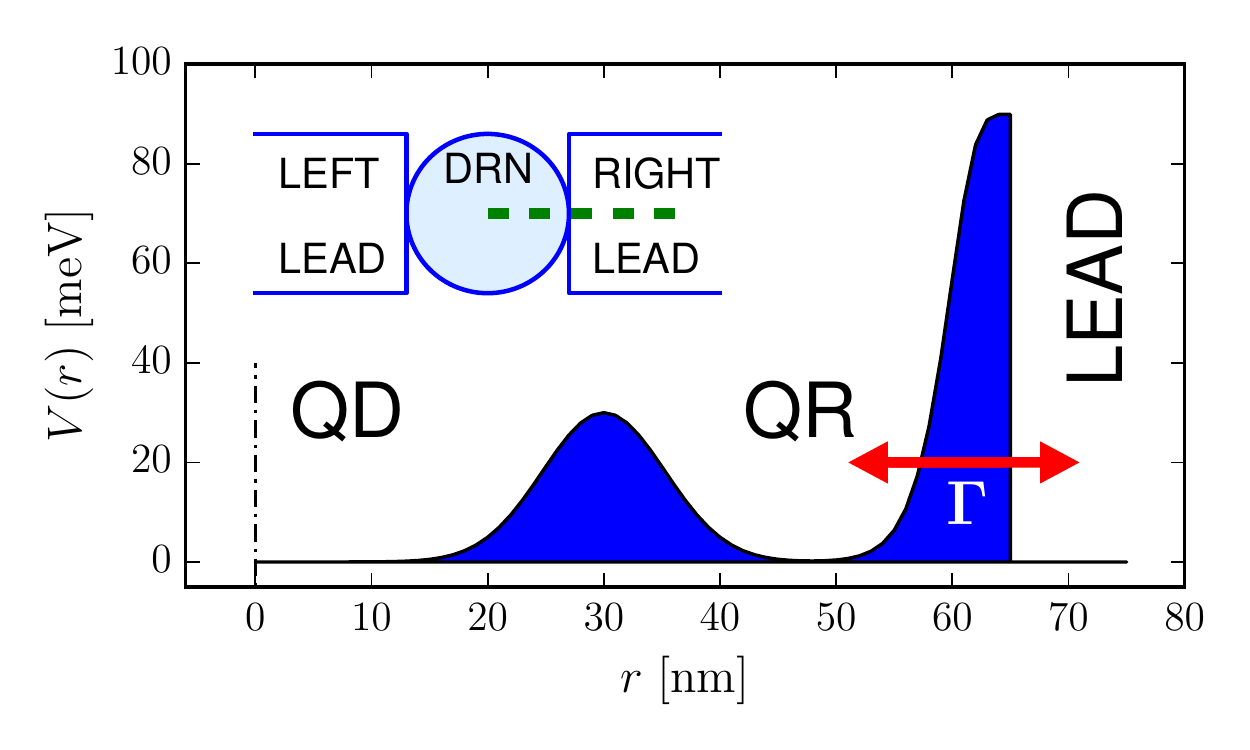}}
  \caption{The cross section of the confining potential with a weakly attached lead for
    $V_{\rm QD}=0$. The red arrow illustrates the tunneling between the DRN and the lead. This
    setup corresponds to the situation presented in Fig. \ref{kwant-scheme}a.
    The thick dashed green line in the inset shows the position of the cross section.
  \label{fig37}} 
\end{figure}

In Fig. \ref{fig36} we show a comparison of the tunneling rate $\Gamma$ between a lead and the
ground state calculated with the help of the Bardeen approach [Eqs. (\ref{ttt}) and
(\ref{gamma_eq})] and determined from the width of the conductance peak. 
One can see there that as far as $V_{\rm QD}$ is small, the ground state wave function is located
in the QD part of the DRN (situation presented in Fig. \ref{vqd}a) and there is no coupling to
the leads. When the bottom of the QD potential is sufficiently high the wave function moves over
towards the QR part (Fig. \ref{vqd}b) and the coupling increases. For $V_{\rm QD}\approx 5$ meV the
wave function is entirely in the QR and further increase of $V_{\rm QD}$ does not affect its shape
(Fig. \ref{vqd}c). As a result $\Gamma$ saturates. In the case presented in
Fig.~\ref{fig36} the leads where attached at a finite distance from the DRN (see
Fig.~\ref{kwant-scheme}a) so the height of the
barrier between the nanosystem and the lead was almost $V_1$. Fig. \ref{fig37} shows
the cross section of the confining potential along the $x$ axis.
For such a high barrier the coupling is weak independently of the distribution of the
wave function. Fig. \ref{fig36} shows that even if $V_{\rm QD}$ is so large that the wave
function is located only in the outer (QR) part of the DRN, $\Gamma$ calculated within
the framework of the Bardeen approach remains in a good agreement with the tight binding model
that takes into account modifications of the wave function shape due to attached leads.

\begin{figure}[htb]
  \centerline{\includegraphics[width=0.5\textwidth]{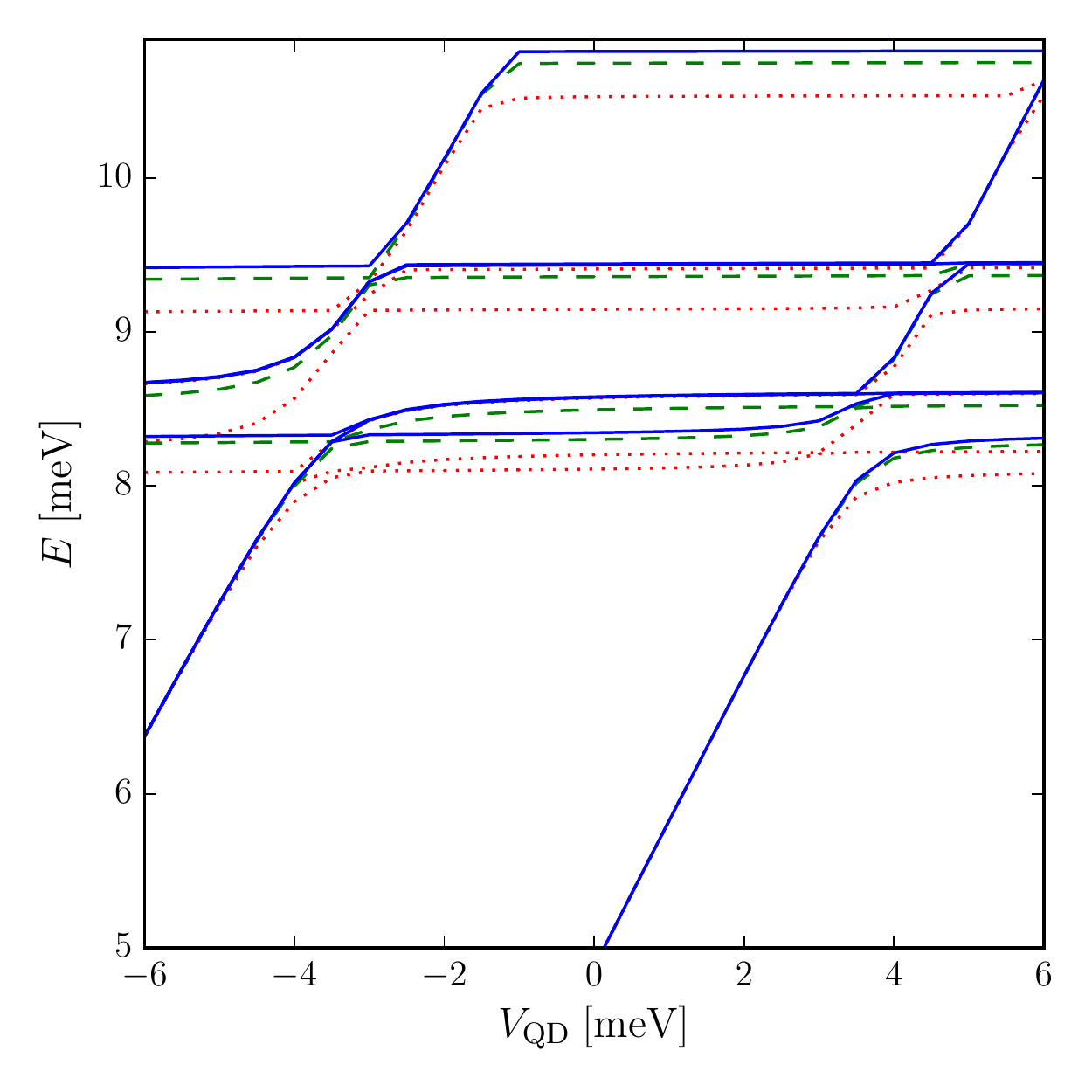}}
  \caption{The energy spectrum as a function of $V_{\rm QD}$ for different couplings to
    the leads. The blue solid, green dashed and red dotted lines show subsequent energies 
    calculated for leads located 65 nm, 59 nm, and 57 nm from the center of the DRN, respectively.
    \label{fig38}}
\end{figure}

\begin{figure}[htb]
  \centerline{\includegraphics[width=0.5\textwidth]{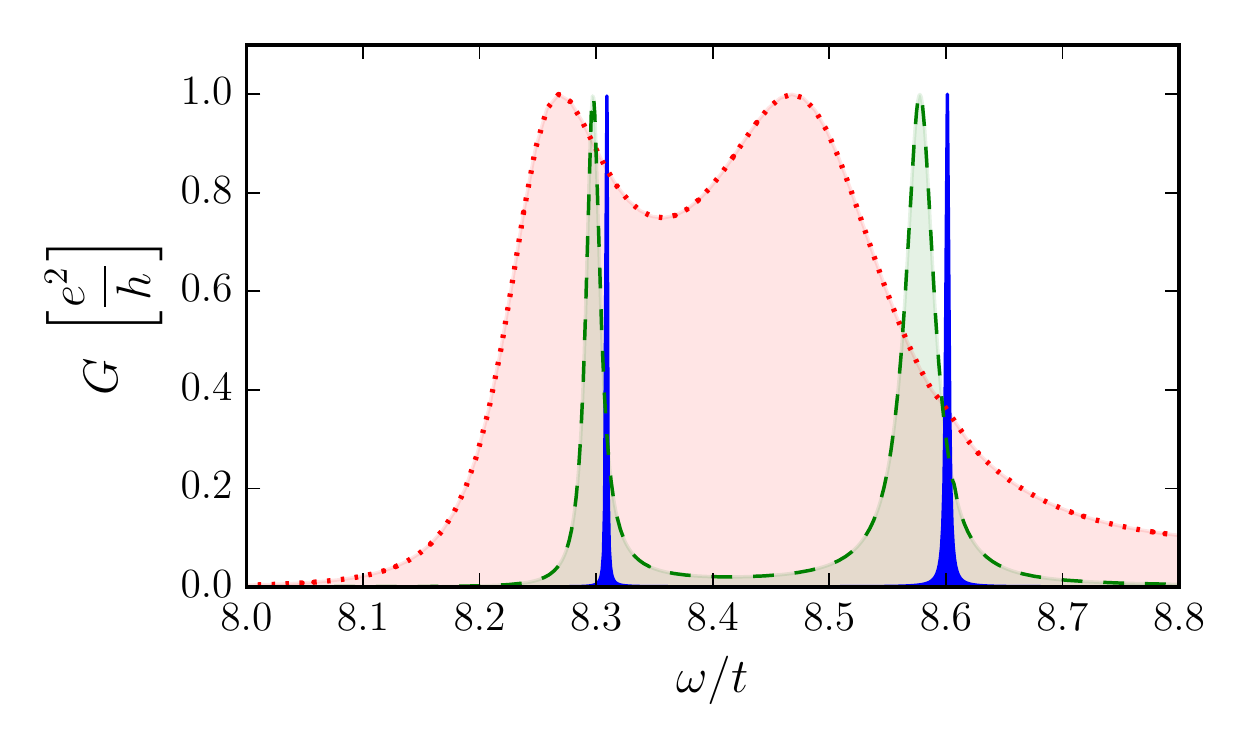}}
  \caption{Conducting peaks for the ground and the first excited state for different
    positions of the leads. The blue solid, green dashed and red dotted lines correspond to leads
    located 65 nm, 63 nm, and 61 nm from the center of the DRN, respectively.\label{fig39}}
\end{figure}

However, as the distance between the lead and the DRN decreases, the tunneling barrier
becomes smaller and the DRN energy spectrum and shape of the wave functions change.
Fig. \ref{fig38} presents how the spectrum changes when the lead is shifted
towards the DRN.
The energy levels in some regimes linearly increase with increasing $V_{\rm QD}$ and in
others are almost constant. Since $V_{\rm QD}$ describes the confining potential only
in the QD part of the DRN, whereas its shape in the QR parts is constant, states with
energies that depend on $V_{\rm QD}$ are located mostly in the central QD. Fig. \ref{fig38}
shows that the energies of these states are not affected by attaching leads, even if
the coupling is relatively strong. On the other hand, the states with energies independent
of $V_{\rm QD}$ are located in the QR, where their matrix elements with states in the
leads are finite and increase with increasing coupling. As a result their energies
do depend on the position of the leads. As shown in Fig. \ref{coupling} not only the
energies, but also the shapes of the wave functions are modified in this regime.
Both the effects, i.e., the reduction of the tunnel barrier between the DRN and the leads
together with the change of the distribution of the wave function, lead to a dramatic increase
of the coupling between the states in the DRN and in the leads. The resulting change
of the tunneling rates is presented in Fig. \ref{fig39}.

\section{Summary}

Combined quantum structures are highly relevant to new technologies in which the control and
manipulations of electron spin and wave functions play an important role. In contrast to real
atoms, DRNs allow for flexible control over the confinement potential which gives rise to wave
function engineering. We have shown that the peculiar structure of a DRN allows us to use an
electrical gating to manipulate the confinement potential so that the coupling between states
in the DRN and in the leads can be changed by orders of magnitude. As a result, by applying the
gate voltage one can easily control the conductance of a DRN. The difference between the
``traditional'' single electron transistor \cite{SIT} and the DRN is that in our case the 
Coulomb blockade is not used. Instead, the gate voltage is used to switch it by 
coupling/decoupling of the DRN to/from the leads \cite{IJZ}.

We have shown that in the case of weakly coupled leads the approximate cylindrical symmetry of
the confining potential can be utilized and the Schr\"odinger equation for only the radial part
of the wave function can be solved numerically. Then, the Bardeen method can be used to calculate
microscopically the geometry--dependent transport properties. However, when the coupling to leads
is stronger, the cylindrical symmetry of the wave functions is broken and other computational
methods must be used. In this paper we have demonstrated that one of the possibilities is package
Kwant, which allows one to do the required calculations relatively easily.

In this paper we have presented the case of sequential tunneling with only a single state in
the bias window. From Fig. \ref{fig38} one can infer that for some shapes of the confining
potential energy levels are very close and such an assumption cannot be fulfilled. When more
than one state is in the bias window and all the states are coupled to the leads, charge
transport can take place through many states. Then, even if we still assume a single electron
transport (the Coulomb blockade does not allow for more than one occupied state to be in the bias
window), one has to take into consideration the relaxation processes between different states.
Such situation has been analyzed in Ref. \onlinecite{MK2}. The situation becomes much more
complicated when more than one electron can occupy the DRN. In this case the Coulomb
interaction and the spin degrees of freedom come into play, what gives a possibility
to control other properties of the nanosystem. Then, a DRN may turn out to be useful also
in spintronics. The work along this line is in progress and the results will be published elsewhere \cite{krakow}.
 
\section*{Acknowledgment}
This work was supported by the National Science Centre (NCN) grant DEC-2013/11/B/ST3/00824.


\begin{thebibliography}{99}

\bibitem{hans} R. Hanson, L. P. Kouwenhoven, J. R. Petta, S. Tarucha, and L. M. K. Vandersypen, Rev. Mod. Phys. {\bf 79}, 1217 (2007).
\bibitem{zipper} E. Zipper, M. Kurpas,  and M. M. Ma\'ska, New J. of Phys. {\bf 14}, 093029 (2012).
\bibitem{MK0} M. Kurpas, B. K\k{e}dzierska, I. Janus-Zygmunt, M. M. Ma\'ska, and E. Zipper, Acta Phys. Pol A {\bf 126}, 20 (2014). 
\bibitem{MK1} M. Kurpas, E. Zipper, and M. M. Ma\'{s}ka, "Engineering of Electron States and Spin Relaxation in Quantum Rings and Quantum Dot-Ring Nanostructures" in "Physics of Quantum Rings", Vladimir M. Fomin (editor), Springer 2014, p. 455
\bibitem{MK2} M. Kurpas, B. K\k{e}dzierska, I. Janus-Zygmunt, A. Gorczyca-Goraj, E. Wach, E.~Zipper, M. M. Ma\'{s}ka, J. Phys.: Condens. Matter {\bf 27}, 265801 (2015).
\bibitem{zkm} E. Zipper, M. Kurpas, J. Sadowski, and M. M. Ma\'ska, J. of Physics: Condensed Matter {\bf 23}, 115302 (2011).
\bibitem{Zeng} Z. Zeng, C. S. Garoufalis, and S. Baskoutas, Phys. Lett. A {\bf 378}, 2713 (2014).
\bibitem{Bardeen} J. Bardeen  Phys. Rev. Lett. {\bf 6}, 57 (1961).
\bibitem{Tersoff} J. Tersoff and D. R. Hamann, Phys. Rev. Lett. {\bf 50}, 1998 (1983).
\bibitem{Tersoff1} J. Tersoff and D. R. Hamann, Phys. Rev. B {\bf 31}, 805 (1985).
\bibitem{Chen} C. J. Chen, Phys. Rev. B {\bf 42}, 8841 (1990).
\bibitem{Chen1} C. J. Chen, Phys. Rev. Lett. {\bf 65}, 448 (1990).
\bibitem{Boykin} T. B. Boykin and G. Klimeck, Eur. J. Phys. {\bf 25}, 503 (2004).
\bibitem{Kwant} Ch. W. Groth, M. Wimmer, A. R. Akhmerov, and X. Waintal, New J. Phys. {\bf 16}, 063065 (2014).
\bibitem{peeters} B. Szafran, F. M. Peeters, S. Bednarek, Phys. Rev. B {\bf 70}, 125310 (2004).
\bibitem{somaschini} C. Somaschini, S. Bietti, N. Koguchi, and S. Sanguinetti, Nanotechnology {\bf 22}, 185602 (2011).
\bibitem{somaschini1} C. Somaschini, S. Bietti, N. Koguchi, and S. Sanguinetti, Appl. Phys. Lett. 
{\bf 97} 203109 (2010).
\bibitem{numerov} B. V. Numerov, Monthly Notices of the Royal Astronomical Society {\bf 84}, 592 (1924);
  B. V. Numerov, Astronomische Nachrichten {\bf 230}, 359 (1927).
\bibitem{BW} A. D. Stone and P. A. Lee, Phys. Rev. Lett. 54, 1196 (1985).
\bibitem{BW1} M. B\"uttiker, Phys. Rev. B {\bf 33}, 3020 (1986). 
\bibitem{BW2} M. B\"uttiker, IBM J. Res. Dev. {\bf 32}, 63 (1988).
\bibitem{SIT} M. A. Kastner Rev. Mod. Phys. {\bf 64} 849 (1992).
\bibitem{IJZ} I. Janus-Zygmunt, B. K\k{e}dzierska, A. Gorczyca-Goraj, M. Kurpas, M. M. Ma\'ska,
and E. Zipper, Acta Phys. Pol. A {\bf 126}, 1171 (2014).
\bibitem{krakow} A. Biborski, A. P. K\k{a}dzielawa, A. Gorczyca-Goraj, E. Zipper, M. M. Ma\'ska, and J. Spa{\l}ek, {\em to be published}.
\end{thebibliography}
\end{document}